\documentclass{iopjournal}
\usepackage{siunitx} % needs to be loaded before physics
\usepackage{physics}
\usepackage{floatrow} 
\usepackage{tipa}
\usepackage{cleveref}
\usepackage{ragged2e}

\newcommand{\kindex}[2]{\ensuremath{{#1}_{\scalebox{0.65}{#2}}}}

\DeclareSIUnit\BL{\textit{L}}
\DeclareSIUnit\BLs{\BL\per\second}
\DeclareSIUnit\bar{bar}
\DeclareSIUnit\inch{"}
\AtBeginDocument{\RenewCommandCopy\qty\SI}
\newcommand{\supfig}[1]{supporting fig.~S{#1}}

\graphicspath{{../figurematter/latex/figs/}}

\renewcommand{\headrulewidth}{0pt}
\renewcommand{\footrulewidth}{0pt}

\begin{document}
\justifying

% \articletype{Paper} %	 e.g. Paper, Letter, Topical Review...

%%%% Article title to be placed here
%% Title suggestions
% - Why do fish avoid swimming at the surface?
% - Rolling and surface effects decrease swimming performance for an undulatory robot
% - From the surface to submerged: undulatory swimming characterisation using a bio-inspired robot
% - The Cost of Constraint: How Tethering and Surface Swimming Affect Robotic Undulatory Performance
% On the cost of constraints: how tethering and depth of submergence affect the swimming performance of  undulatory robots
% - Experimental Constraints in Robotic Undulatory Swimming: Tethering Effects and Depth Dependence
% - The Effects of Tethering and Depth on Robotic Undulatory Swimming
% - Free versus Tethered Undulatory Swimming: Effects of Mechanical Constraints and Submergence Depth on Robotic Performance

\title{Tethering and depth of submergence affect the swimming performance of undulatory robots}

\author{%%%% Author details
Alexandros Anastasiadis$^{1,2}$,  Auke J. Ijspeert$^2$ and Karen Mulleners$^{1,*}$}

%%%%%%%%% Insert author address here
\affil{$^1$Unsteady Flow Diagnostics Laboratory, Institute of Mechanical Engineering, École Polytechnique Fédérale de Lausanne (EPFL), Lausanne, Switzerland}

\affil{$^2$Biorobotics Laboratory, Institute of Bioengineering, École Polytechnique Fédérale de Lausanne (EPFL), Lausanne, Switzerland}

\affil{$^*$Author to whom any correspondence should be addressed.}

%%%% Keyword entries to be placed here %%%%

%%%% Insert corresponding author and its email address}
\email{karen.mulleners@epfl.ch}

%%%% Abstract text to be placed here %%%%%%%%%%%%
% Should not normally be more than 300 words - 260 now
\begin{abstract} 
\justifying
Over the past few decades, biomimetic robotic experiments have significantly advanced our understanding of undulatory swimming. Compared to animal experiments, robotic experiments offer repeatability and controlled parameter variations, but the robots operate under constraints that differ from those experienced by their natural counterparts. Freely swimming robots often remain on the surface, whereas most undulatory fish, including eels, are typically fully submerged during locomotion. Studies focusing on submerged swimming commonly rely on tethered robots to maintain depth control. 
This study examines the performance implications for free versus tethered swimming at the surface, and for tethered swimming at the surface to tethered fully submerged swimming, using the robotic undulatory swimmer 1-guilla.
The robot was tested in two configurations: free swimming in a pool and tethered swimming in a water channel at the surface and at varying depths down to three body heights. We varied kinematic input parameters and quantified performance in terms of swimming speed, cost of transport, and body kinematics. Our results reveal that at the surface, tethered swimming achieves speeds comparable to free swimming but at a lower energetic cost. 
This reduction in cost of transport is attributed to the mechanical stabilisation imposed by tethering, including the suppression of body roll and lateral head motion.
Increasing submergence depth improved both the maximum speed and energy efficiency by more than 10\% relative to the surface swimming performance. As the body kinematics remained unchanged when submerged, the performance deficit near the surface is attributed to increased wave drag. Overall, our findings provide explanations and insights into discrepancies in results obtained for tethered and free-swimming robotic studies, they highlight the hydrodynamic challenges of surface locomotion, and can help explain why natural undulatory swimmers predominantly favour submerged propulsion.
\end{abstract}
%%%%%%%%%%%%%%%%%%%%%%%%%%%

\section{Introduction}
%% Biomimicry in Aquatic Locomotion
Biomimicry is the study of replicating natural strategies to address technical challenges \cite{Bhushan2009}. 
Over the past few decades, biomimicry has become a powerful framework in the field of robotics and fluid mechanics, particularly for studying aquatic locomotion \cite{Cui2023}. 
In aquatic locomotion, biomimetic research seeks to understand and reproduce swimming strategies and associated body kinematics observed across a wide range of species, from microorganisms \cite{Huang2016} to large marine animals \cite{Triantafyllou1995}. 
Among the various modes of swimming, undulatory propulsion is one of the most widespread and effective strategies in nature, adapted across diverse species and environments \cite{Santo2021}.
Undulatory swimming animals generate thrust by wave-like body displacements that travel along the body from head to tail \cite{Gray1933,Smits2019}.

The motivation for biomimetic research on undulatory locomotion extends beyond technological curiosity. 
Natural undulatory swimmers have evolved solutions that show remarkable performance in speed \cite{Bainbridge1958}, acceleration \cite{Webb1976}, efficiency \cite{Lighthill1970,VanGinneken2005}, and maneuverability \cite{Epps2007,Triantafyllou2016}.
By understanding the underlying principles of biological swimming, we can develop more efficient underwater vehicles, advance our fundamental understanding of fluid-structure interactions, and gain insights into the processes that shaped evolution \cite{Triantafyllou2000,Wainwright2020,Gravish2018}. 
These insights are increasingly relevant as demand grows for autonomous underwater vehicles in applications such as ocean exploration, environmental and water-quality monitoring, and marine infrastructure inspection \cite{Wu2017,Fish2020,Sverdrup-Thygeson2016,Bayat2017,Katzschmann2018}.

%% The Role of Robotic Experiments
% Why do researchers build robotic platforms instead of simply studying live animals? 
Compared to living organisms, robotic set-ups offer experimental advantages such as repeatability and precise parametric control.
Relevant parameters of interest for undulatory swimming include body stiffness \cite{Zhong2021}, shape \cite{King2023}, undulation amplitude, frequency, and wavelength \cite{Zhu2019, Nangia2017}.
In animals, multiple parameters may be coupled or vary simultaneously \cite{Gazzola2014,Gemmell2015}.
In robots, individual parameters can be systematically varied while other factors are forced to remain constant.  
This dedicated parametric control enables us to isolate specific variables and establish relationships that would be challenging to discern from observations of swimming animals \cite{Anastasiadis2023}.
Furthermore, we can make robots perform motions that are not observed in animals to aid identifying whether the performance of fish is indeed as optimal as is often suggested \cite{Lauder2007}.
Finally, there are less ethical and technical constraints to equip a robot with sensors than to extract sensorial feedback from live animals.
The sensorial feedback can be used for closed-loop control and neuromechanical modelling applications \cite{Thandiackal2021, Ramdya2023}.

Robotic platforms also facilitate flow visualisation and force measurements that are challenging or impossible to do with live animals. 
Combined measurements of the complex flow structures generated by robotic swimmers by means of time-resolved particle image velocimetry (PIV) \cite{Leftwich2012,Quinn2015,Zhong2021,Berlinger2021}, hydrodynamic forces \cite{Curet2011,Sanchez-Rodriguez2021}, and kinematic tracking of body kinematics \cite{White2021} are desirable to reveal how body kinematics translate into thrust production and propulsive efficiency \cite{Zhong2021}.

Yet, to obtain the desired parameter control and measurement capabilities on the experimental platforms, robots are often operated under constraints that differ from those experienced by their natural counterparts. 
Freely swimming robots often remain on the surface \cite{Yan2008,Manfredi2013,Porez2014,Bayat2016,Li2020,Li2021,Anastasiadis2023}, whereas most undulatory fish, including eels, typically avoid swimming near the surface \cite{Videler1993}.
Robots that swim at the surface are easier to visually track and do not need to be neutrally buoyant, which simplifies their design.  
Studies focusing on submerged swimming commonly rely on tethered configurations where the robot is fixed in place by a shaft or rod, typically operating in a water channel or towing tank \cite{Triantafyllou1995,Hultmark2010,Curet2011,Salumae2013,Chambers2014,Bayat2016,Jusufi2017,Nangia2017,Gibouin2018,Zheng2018,Zhong2021,Sanchez-Rodriguez2021,Obayashi2026}.
The main advantages of tethered configurations include precise control of the swimming depth, direct measurement of hydrodynamic forces using load cells attached to the tether beam, and automated experimental protocols \cite{mulleners.2024}. 
Furthermore, a fixed location of the robot facilitates flow measurements. 
But, tethering alters the swimming dynamics by reducing the degrees of freedom of the motion. 
In particular, the rotational body motion around the roll and the pitch axes, the vertical (heave) and lateral translation (sway) are often constrained by tethering the robot.  
In some experimental configurations, bearings or cord mounts are introduced to restore selected degrees of freedom required for specific measurements \cite{Zhu2019}.
In the present tethered configuration, all translational degrees of freedom are fixed, roll and pitch are constrained, and bearings permit yaw rotation about the vertical shaft.
By contrast, the free-swimming robot is not mechanically constrained in these degrees of freedom, but buoyancy and interaction with the free surface passively limit its heave and pitch motions.

%%%%% Our objectives
This study addresses two important methodological questions that frequently arise in biomimetic robotic undulatory swimming experiments.
First, what are the performance differences between free-swimming and tethered configurations?
Second, how do surface versus fully submerged swimming conditions affect the experimental outcomes?
%%%%% Our methodology
To address these questions, we conduct systematic experiments with our bio-inspired undulatory robot, 1-guilla, in two different configurations: free swimming and tethered swimming at variable depths. 
The robot swimming performance is evaluated through measurements of swimming speed, cost of transport, time-resolved forces, and tracking of the body kinematics. 
We first compare results of the robot swimming at the surface for tethered and free-swimming configurations. 
Roll motion and lateral head amplitude during surface swimming are quantified to investigate their causal effect on swimming performance losses. 
We then explore the effects of the submersion depth across the full range, from surface swimming to completely submerged swimming, using the tethered configuration.  
Free submerged swimming was not performed due to practical limitations related to buoyancy control and non-uniform mass distribution along the body.  

\begin{figure}[!t]
\centering\includegraphics{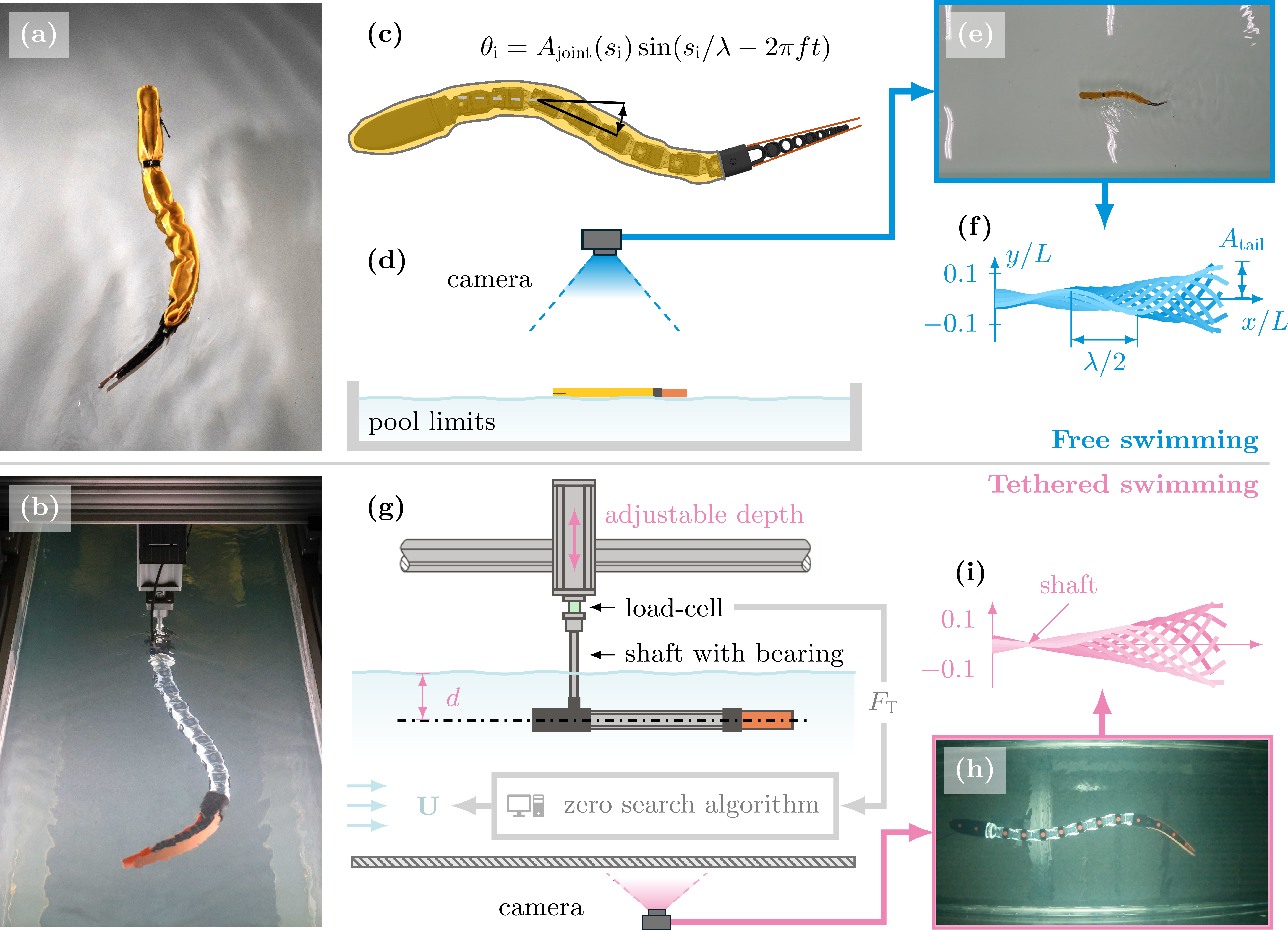}
\caption{Overview of the experimental set-up for free swimming (light blue colour scheme) and tethered swimming configurations (pink colour schemes).
\textbf{(a)} Photograph of the free-swimming version of the undulatory robot 1-guilla in the experimental pool.
\textbf{(b)} Photograph of the tethered swimming version of 1-guilla in the water channel.
\textbf{(c)} Schematic of the robotic multi-joint configuration. Both robots consist of a chain of 8 motors and an external flexible tail attached outside the waterproof suit.
\textbf{(d)} Schematic overview of experimental set-up for free swimming in the pool. 
A camera, placed above the pool, records videos of the performed swimming motions.
\textbf{(e)} Example frame from the video recordings in the pool.
\textbf{(f)} Extracted midline evolution of the robot and projected on a single-axis Cartesian coordinate system. 
The performed tail amplitude (\kindex{A}{tail}) and wavelength ($\lambda$) are calculated from the projected midlines.
\textbf{(g)} Schematic overview of experimental set-up in the water channel. 
The robot is held by a shaft with bearings, depth adjustment beam, and a load cell that records all components of the force and torque.
The recorded thrust force from the load cell (\kindex{F}{T}) is fed into a zero search algorithm that adjusts the flow speed to achieve zero net thrust.
Once the algorithm converges, a camera located below the test section records the swimming motion.
\textbf{(h)} Example frame from the video recordings in the water channel.
\textbf{(i)} Extracted midline evolution of the robot for the tethered swimming configuration. 
The tethering of the shaft is captured as a node in the recorded midline evolution.
}
\label{fig:expsetup}
\end{figure}

\section{Methodology}
Two types of experiments were conducted for this study: free swimming experiments in a pool and tethered swimming experiments in a water channel (\cref{fig:expsetup}). 
Two geometrically similar versions of the bioinspired robot 1-guilla were developed, each tailored to the specific requirements of the two experimental setups. 

\subsection{Robot design}
The robot 1-guilla is inspired by and modelled after the morphology of eels. 
Both robot versions measure $L = \qty{0.85}{\meter}$ in length and consist of a head, eight actuated body segments, and a flexible tail. 
The body segments are linked by custom 3D-printed rigid parts. 
Each segment is actuated by a Dynamixel XM430-W210-R servo motor. 
The passive, flexible tail is magnetically attached to the last segment (\cref{fig:expsetup}a,b).

The free-swimming robot version (\cref{fig:expsetup}a) floats on the water surface and operates wirelessly. 
All motion-control electronics are housed in the robot head.
A Raspberry Pi Zero 2 W microcontroller sends the motor commands via a Dynamixel U2D2 interface in position control mode, with a control timestep of \qty{16}{\milli\second}. 
The board and the motors are powered by an \qty{11.1}{\volt}, \qty{1500}{\milli\ampere\hour} Li-Po battery.
An external computer initiates and terminates the motion via Wi-Fi.
A waterproof suit covers the head and motor components. 
The suit is made of yellow rip-stop fabric coated with a thermo-adhesive material. 
The thermo-adhesive material is used to seal the suit seams, bonding the parts together under heat and high pressure in a heat press. 
A waterproof zipper (TIZIP, Master-Seal 10, pressure proof) is added to easily open and close the suit to exchange the batteries.
When the zipper is closed, air is sealed within the suit, which causes the robot to float at the water surface.

The tethered swimming robot version (\cref{fig:expsetup}b) was redesigned to be attached to a shaft in the recirculating water channel. 
To allow for repeated and continuous experiments, the robot is controlled by an external computational and power unit. 
The computational unit consists of a Raspberry Pi 5 board interfaced with a Dynamixel U2D2 unit, using the same software as in the free-swimming version. 
Power is supplied by a DC power source (\qty{12}{\volt}, maximum \qty{3}{\ampere}).
The robot is rigidly attached to a vertical shaft, which fixes all translational degrees of freedom and constrains rotation in roll and pitch.
The shaft is mounted on bearings that permit yaw rotation about the vertical axis.
The shaft is connected to a multi-axis load cell (ATI Mini45), enabling direct measurement of the forces and torques generated during swimming.

This version of the robot incorporates a waterproofing design tailored to long-duration tethered use \cite{Liu2025}. 
The waterproofing is achieved by a transparent thermoplastic polyurethane (TPU) mountain bike inner tube (Schwalbe Aerothan $\qty{27.5}{\inch}$ MTB+) that covers the motors. 
The tube is secured and sealed at the head by an adjustable circular pipe clamp.
At the tail, the tube is closed by heat sealing. 
To ensure that the modified waterproofing design does not bias performance measurements, a direct comparison between the two robot versions was conducted under identical tethered conditions in the water channel (\supfig{1}).
The results show comparable stride length and cost of transport for identical input kinematics, indicating that the waterproofing and structural differences have a negligible effect on the measured performance.
Power and communication cables exit the rigid head through a waterproof connector. 
Due to reduced air entrapment, this version is negatively buoyant, resulting in a net downward vertical force of \qty{2.61}{\newton} when submerged in water.

\subsection{Input kinematics}
The input joint kinematics of the robot followed a travelling wave of increasing amplitude, applied in position control (\cref{fig:expsetup}c):
\begin{align}\label{eq:theta}
\kindex{\theta}{i}(t)=\kindex{A}{joint}(\kindex{s}{i}) \sin{(\kindex{s}{i}/\kindex{\lambda}{input}-2\pi ft)},
\end{align}
where $\kindex{\theta}{i}(t)$ is the angle control input for the $i$-th joint, with $i=1$ the motor closest to the head, $\kindex{s}{i}$ is the length-normalised position of the $i$-th motor joint, $\kindex{A}{joint}$ is the input joint angle amplitude as a function of $\kindex{s}{i}$, $f$ is the undulation frequency, kept constant at \qty{1}{\hertz}, and $\kindex{\lambda}{input}$ is the wavelength.
The amplitude growth along the body was defined by a polynomial derived from observations of American eels \cite{Tytell2004a,Anastasiadis2024}:
$\kindex{A}{joint}(\kindex{s}{i})=(1+0.323(\kindex{s}{i}-1)+0.310(\kindex{s}{i}^2-1))
\kindex{A}{joint,max}$ , where $\kindex{A}{joint,max}$ is the maximum joint angle amplitude applied at the tail joint.

The input wavelength and maximum joint amplitude were systematically varied in both experimental setups.
The input wavelength $\kindex{\lambda}{input}/L$, ranged from \numrange{0.5}{1} in steps of $1/14$.
The maximum joint amplitude $\kindex{A}{joint,max}$ ranged from \qtyrange{10}{30}{\degree} in the tethered-swimming case and the upper limit was increased up to \ang{45} in the free-swimming case.
The step size was \ang{5} in both cases. 

\subsection{Free swimming experiments}
The free swimming experiments were conducted in the swimming pool of \qtyproduct{6x2x0.3}{\meter} of the BioRob laboratory at EPFL (\cref{fig:expsetup}d). 
In each trial, the robot was released from one end of the pool and swam along the length of the pool. 
Experiments were repeated five times ($N=5$) for each input parameter combination to account for statistical variability and deviations from straight-line trajectories. 
The motion of the robot was captured by a GoPro Hero 8 Black camera.
The camera was mounted above the center of the pool at a height of \SI{2.25}{\meter} from the water surface.
Videos were captured at $30$ frames per second with a resolution of \qtyproduct{1920x1080}{pixel} (\cref{fig:expsetup}e). 

%%%%
The midline of the robot was extracted from video frames using an image-processing and clustering algorithm (see Supplementary Information and \cite{Raynaud2025} for more details). 
The algorithm robustly captures both small and large body curvatures and ensures constant body length, providing reliable midline coordinates used for analysis.

%%%%
The speed of the robot was quantified from the extracted midline coordinates and the trajectory of the motion. 
The trajectory of the motion was obtained as a second-degree polynomial fit to the cloud of midline points across all frames in the global Cartesian reference frame.
The global midline coordinates $(\kindex{x}{global}, \kindex{y}{global})$ were projected onto the trajectory to yield a single-axis representation of the midline $(x, y)$ (\cref{fig:expsetup}f). 
The longitudinal coordinate ($x$) was defined as the curvilinear distance along the fitted trajectory between the projection of each midline point onto the trajectory and the projection of the head point.
The lateral projected component $y$ was defined as the orthogonal distance between each midline point and the trajectory at each time step, representing the lateral displacement relative to the swimming direction.
The time-resolved speed $U(t)$ was obtained as the temporal derivative of the projection of the head trace along the swimming trajectory.  
The initial release transient was excluded by discarding the first \qty{30}{\percent} of each trial.
The remaining signal covered five undulation periods and was phase averaged. 
The steady-state swimming speed $\overline{U}$ was then calculated as the mean of the phase-averaged velocity signal.
The variability across the analysed periods is low, indicating that high-frequency tracking noise has limited influence on the calculated mean speed.

%%%%
The rolling angle of the robot ($\gamma$) during swimming was measured with an IMU sensor. 
The sensor (LSM6DSO) was mounted on the head segment and communicated with the Raspberry Pi board via an $I2C$ bus. 
Measurements were recorded at each time step of the motion-control algorithm, providing the three-axis acceleration of the head.

The rolling angle $\gamma$ was estimated from these accelerations as the projection of the gravitational acceleration, according to \cite{Luczak2017}:
\begin{align}\label{eq:gamma}
\gamma = \sin^{-1}\left(\frac{\kindex{\alpha}{y}}{\sqrt{\kindex{\alpha}{x}^2+\kindex{\alpha}{y}^2+\kindex{\alpha}{z}^2}}\right)\ ,
\end{align}
where $\kindex{\alpha}{x}$, $\kindex{\alpha}{y}$, and $\kindex{\alpha}{z}$ are the components of acceleration in the head-attached coordinate system, with the $x$-axis aligned with the longitudinal axis of the robot, the $y$-axis aligned with the lateral axis, and the $z$-axis aligned with the dorsoventral axis of the robot.

\subsection{Tethered swimming experiments \& set-up automation}
The tethered swimming experiments were conducted in the \qtyproduct{3x0.6x0.6}{\meter} test section of the recirculating water channel of the UNFoLD laboratory at EPFL (\cref{fig:expsetup}g). 
The head of the robot was attached to a shaft at \qty{10.2}{\centi\meter} from the front edge of the robot.
The attachment point corresponds to $0.12 L$, a location in the undulation of swimming animals that often exhibits minimal lateral amplitude \cite{Santo2021}. 
The shaft was mounted on two ball bearings, which allowed low-friction rotation around its axis.
The outer housing of the bearings is attached to a six-component load-cell (ATI mini45).
The forces and torques were recorded by the load-cell at a frequency of \qty{1000}{\hertz}. 
A zeroing procedure for the load-cell was performed before each trial, and at regular intervals to subtract baseline forces and torques measured when the robot was attached without flow or motion.

The swimming depth $d$ of the robot was controlled through an adjustable attachment on the shaft (\cref{fig:expsetup}g).
All tested depths were normalised by the robot height $h = \qty{4.5}{\centi\meter}$ and measured with respect to the robot mid-height.
Surface swimming corresponds to $d/h = 0$.
Partial submersion corresponds to $d/h = 0.5$, with the top of the robot touching the water surface.
Full submersion was tested at $d/h = 1.2$, $d/h = 2$, and $d/h = 3$.
The tested depth conditions are summarised in \cref{tab:depths}.

\begin{table}[h!]
\centering
\caption{Tested swimming depths normalised by the robot height $h$.}
\label{tab:depths}
\begin{tabular}{l c}
\hline
Condition & Normalized depth $d/h$ \\
\hline
Surface swimming & $0$ \\
Partial submersion (top at surface) & $0.5$ \\
Full submersion (shallow) & $1.2$ \\
Full submersion (intermediate) & $2.0$ \\
Full submersion (deep) & $3.0$ \\
\hline
\end{tabular}
\end{table}

The motion of the robot was captured with a Raspberry Pi camera mounted below the water channel.
The camera was recording at $30$ frames per second with a resolution of \qtyproduct{1920x1080}{pixel}. 
Image distortion caused by the \SI{4}{\centi\meter} thick plexiglass bottom of the channel was corrected using a custom calibration target with a checkerboard pattern.
More information about the image dewarping and conversion from pixel coordinates to metric coordinates can be found in the \supfig{3}.

The midline of the robot was extracted with the aid of red markers magnetically attached at fixed locations along the body. 
The marker positions were identified by thresholding the RGB channels of the video frames, generating a binary mask of pixels below a specified intensity value. 
The centre of each marker cluster was then selected as a midline point. 
The resulting midline is directly in a single-axis representation.

The swimming speed was determined with a zero-search algorithm that identifies the flow velocity at which the robot exerts zero net thrust. 
The algorithm reads the streamwise force measured by the load cell and adjusts the oncoming flow by controlling the pump rotation. 
The search begins with two test speeds and the corresponding thrust forces. 
The next candidate speed is chosen at the intersection of the thrust-force gradient with the zero-thrust axis. 
The iteration follows the regula falsi method until convergence \cite{Conte2017}.
A convergence threshold of \qty{20}{RPM} of the pump speed, corresponding to $\approx \qty{0.01}{\meter/\second}$, ensured accurate results.
A complete measurement lasted approximately \qtyrange{75}{80}{\second}, including the time required to adjust the pump between successive candidate speeds.
The drag force of the supporting pole was measured independently for all tested depths as a function of the flow speed (\supfig{4}).
A decrease in speed close to the surface due to boundary effects was also measured and characterised (\supfig{5}).
The algorithm considers both the pole drag and the boundary effect to increase the accuracy across the depths tested here. 

\subsection{Power consumption}
The electrical power consumption ($P$) of the motors is recorded for both versions of the robot. 
The instantaneous voltage, $\kindex{V}{i}(t)$, and current, $\kindex{I}{i}(t)$, is obtained for all eight motors, $i=1$ to $i=8$, at each time step. 
The average power consumption for each experiment is calculated considering \num{5} periods of motion: 
\begin{equation}
\overline{P}=\overline{\sum\limits_{i=1}^{8}{
\kindex{V}{i}(t)\kindex{I}{i}(t)}
}.
\end{equation}

\subsection{Emergent swimming characteristics}
The maximum tail amplitude $\kindex{A}{tail}$ (\cref{fig:expsetup}f) was calculated from the single-axis projected kinematics in the Cartesian space for both experimental setups. 
The value of $\kindex{A}{tail}$ was defined as the average over five undulation periods as half the peak-to-peak amplitude of the last body node.

The performed wavelength $\lambda$ (\cref{fig:expsetup}f) was calculated from the absolute curvature of the midline in Cartesian space.
Unlike the input wavelength $\kindex{\lambda}{input}$ defined in \cref{eq:theta}, which is prescribed in the joint-angle space, $\lambda$ is obtained from the measured midline kinematics.
It was defined as the mean over five undulations of twice the longitudinal distance between consecutive curvature peaks.

\section{Results}

\begin{figure}[!t]
\centering\includegraphics{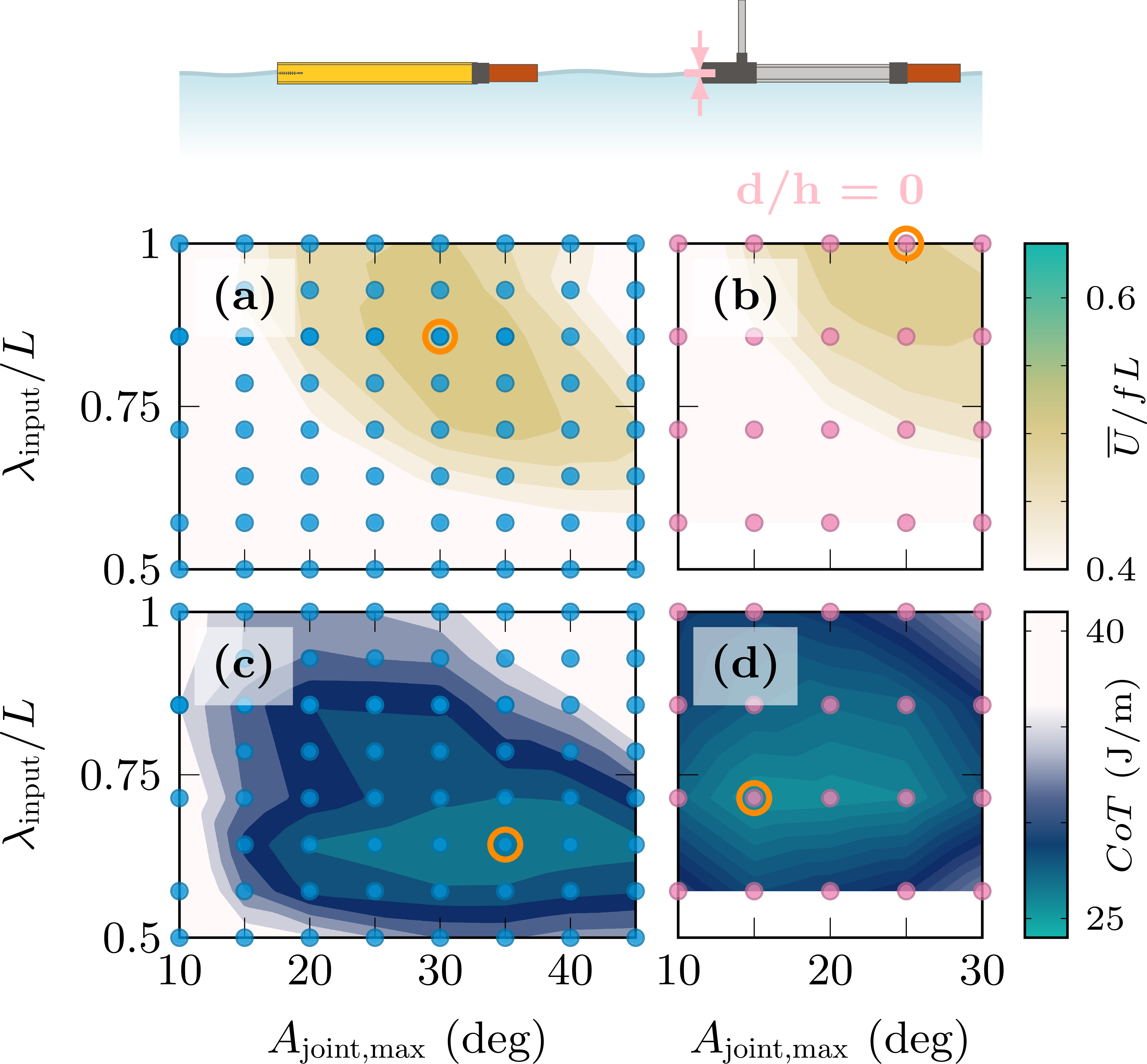}
\caption{Surface swimming performance maps of normalised stride length (a,b) and cost of transport (c,d) for free (left) and tethered (right) swimming as a function of the input kinematic parameters ($\kindex{\lambda}{input}/L$ and $\kindex{A}{joint,max}$). Orange circles indicate the best performance points for every map.}
\label{fig:freevstetheredmaps}
\end{figure}

\subsection{Free versus tethered swimming}

\subsubsection{Performance overview}

\Cref{fig:freevstetheredmaps} shows the performance maps for surface swimming as a function of the input kinematic parameters ($\kindex{\lambda}{input}/L$ and $\kindex{A}{joint,max}$) for both free and tethered configurations. 
Performance is expressed by the normalised stride length ($\overline{U}/fL$) and by the cost of transport ($CoT = \overline{P}/\overline{U}$), which is a measure of swimming efficiency.

\Cref{fig:freevstetheredmaps}a,b show the distribution of stride length over the tested range of input parameters.
For $\kindex{A}{joint,max}<\ang{25}$ and $\kindex{\lambda}{input}/L<0.74$, stride length increases with both $\kindex{A}{joint,max}$ and $\kindex{\lambda}{input}/L$.
An optimum region is observed at higher parameter values, where the normalised stride length approaches the same maximum value $\kindex{\overline{U}/fL}{max,surface}=0.54$ at input parameters of $\kindex{A}{joint,max}=\ang{30}$ and $\kindex{\lambda}{input}/L=0.85$ for free swimming, and $\kindex{A}{joint,max}=\ang{25}$ and $\kindex{\lambda}{input}/L=1$ for tethered swimming.
Joint amplitudes above \ang{40} yield reduced stride lengths.
Different optimal parameter combinations can produce comparable tail motions during tethered and free surface swimming because the interplay between the joint amplitude and wavelength determines the resulting tail excursion. 
The relationship between the tail motion and the resulting normalised stride length is examined in more detail below.
In tethered swimming, amplitudes beyond \ang{30} bring the tail too close to the channel wall and are avoided.
The overall distribution of stride length shows similar trends for free and tethered surface swimming.

\Cref{fig:freevstetheredmaps}c,d show the distribution of the cost of transport over the tested range of input parameters.
For free swimming, the lowest values occur for $\kindex{\lambda}{input}/L$ between \numrange[range-phrase={~and~}]{0.60}{0.75} and for $\kindex{A}{joint,max}$ between \qtyrange[range-phrase={~and~}]{20}{45}{\degree}.
For tethered swimming, the optimum region lies at $\kindex{\lambda}{input}/L \approx \numrange{0.72}{0.75}$ and $\kindex{A}{joint,max} \approx \SIrange{14}{25}{\degree}$.
Values of the input parameters outside these ranges increase the cost of transport.
Free swimming produces values between $CoT = \SIrange[range-phrase={~\text{and}~}]{29.8}{80}{\joule / \meter}$, covering both low and high energetic costs.
Tethered swimming produces a narrower range of values between $CoT = \SIrange[range-phrase={~\text{and}~}]{23.8}{33}{\joule / \meter}$, limiting the cost to a lower and more stable range.
Tethering reduces the occurrence of high costs of transport and makes energy expenditure less sensitive to the input kinematics.
Despite this shift in magnitude, the overall distribution of the cost of transport follows similar patterns in free and tethered surface swimming.

Stride length and cost of transport show consistent performance patterns between free and tethered surface swimming.
Stride length increases with joint amplitude and wavelength up to an optimum region, and decreases at high amplitudes.
Cost of transport reaches a minimum within a restricted range of input parameters of lower $\kindex{\lambda}{input}/L$ and similar to lower $\kindex{A}{joint,max}$, consistent with the previously identified trade-off between speed and efficiency \cite{Anastasiadis2023}. 
Free swimming produces a wider distribution of energetic costs, whereas tethered swimming limits the cost of transport to a narrower and more stable range.
Tethering alters the energetic landscape by suppressing high costs of transport while preserving the overall trends in performance of the cost of transport maps.

\subsubsection{Kinematic insights}

\begin{figure}[!t]
\centering\includegraphics{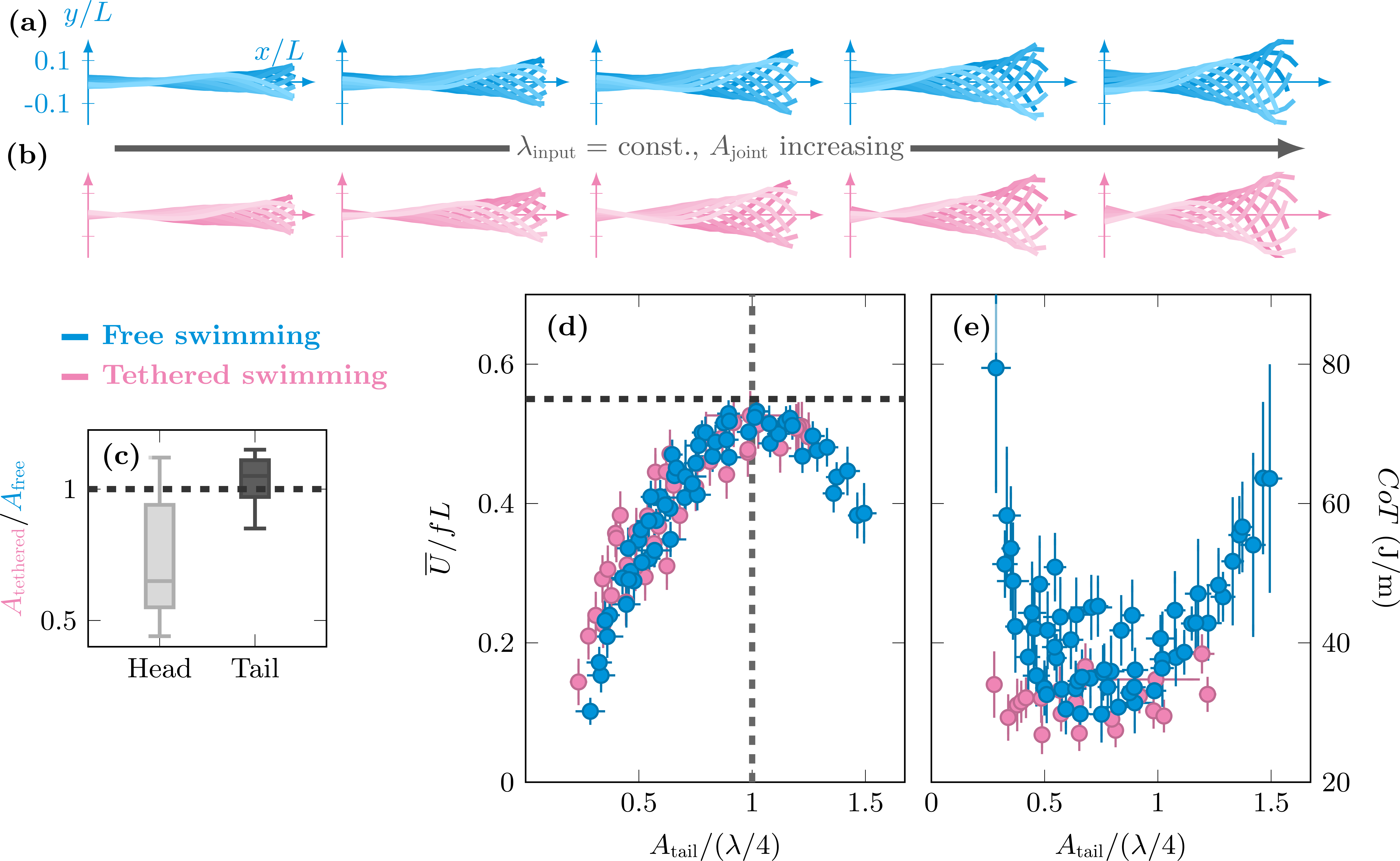}
%% where xxxxxx name represents "figurename.eps"
\caption{
Kinematics analysis of free surface swimming. 
Example midlines for free \textbf{(a)} and tethered \textbf{(b)} swimming during one undulatory period. 
From left to right, \kindex{A}{joint,max} increases from \qtyrange{10}{30}{\degree} in steps of \ang{5}, for $\kindex{\lambda}{input}/L = 0.85$. 
\textbf{(c)} Statistics of the ratio of tethered to free swimming for head and tail amplitudes of the same kinematic inputs.
\textbf{(d)} Normalised stride length as a function of the specific tail amplitude.
\textbf{(e)} Cost of transport as a function of the specific tail amplitude.
}
\label{fig:freevstetheredkin}
\end{figure}

The resulting kinematics are analysed here to understand the origin of the observed performance similarities and differences.
The actual body posture of the robot is a result of both the commanded input parameters $\kindex{\lambda}{input}/L$ and $\kindex{A}{joint,max}$ and the interaction with the surrounding fluid.
We focus here on the distribution of motion along the body, from head to tail (\cref{fig:freevstetheredkin}a,b).

Selected resulting midline kinematics are presented in \cref{fig:freevstetheredkin}a and b for free and tethered swimming, respectively. 
Both configurations are compared for the same input parameters.
The input wavelength is fixed at $\kindex{\lambda}{input}/L = 0.85$, and the joint amplitude increases from $\kindex{A}{joint,max}= \qtyrange{10}{30}{\degree}$, in steps of \SI{5}{\degree} from left to right. 
The resulting lateral amplitude decreases from the head towards a minimum before increasing monotonically toward the tail.
In free swimming, the minimum displacement is non-zero and located at $x/L=\numrange{0.2}{0.3}$.
In tethered swimming, the minimum displacement is zero and imposed by the location of the tether shaft at $x/L=0.12$.
The largest lateral amplitude is always observed at the tip of the tail and remains comparable between the two conditions.
Larger input amplitudes produce larger lateral displacements along the body in both configurations.

The amplitude ratios between the tethered and free swimming configurations are shown in \cref{fig:freevstetheredkin}c.
The tail amplitude ratio $\kindex{A}{tail,tethered}/\kindex{A}{tail,free}$ results in a median value of $\overline{\kindex{A}{tail,tethered}/\kindex{A}{tail,free}}=1.05$ with an interquartile range of \numrange{0.96}{1.11}.
This distribution centred around unity indicates that the performed tail amplitude is similar between the two configurations.
The head amplitude ratio $\kindex{A}{head,tethered}/\kindex{A}{head,free}$ has a lower median value of $\overline{\kindex{A}{head,tethered}/\kindex{A}{head,free}}=0.67$.
Tethering near the head substantially reduces the lateral amplitude close to the attachment point.
Still, the tail motion is largely preserved. 

Stride length is governed by the tail motion \cite{Anastasiadis2023} and is shown here against the specific tail amplitude $\kindex{A}{tail}/(\lambda/4)$ for both free and tethered swimming (\cref{fig:freevstetheredkin}d).
The specific tail amplitude scales the stride length output for free swimming \cite{Anastasiadis2023}. 
In both free and tethered swimming, stride length increases with increasing specific tail amplitude until unity.
For $\kindex{A}{tail}/(\lambda/4)=1$, the normalised stride length reaches a maximum value of \num{0.55}.
Further increase of the specific tail amplitude will cause a decrease in stride length.  
Free and tethered swimming configurations present similar performance, mainly attributed here to the similar tail amplitude.

The cost of transport is shown in \cref{fig:freevstetheredkin}e as a function of the specific tail amplitude for both free and tethered swimming. 
Cost of transport decreases with increasing specific tail amplitude until $\kindex{A}{tail}/(\lambda/4)\approx 0.65$, where it reaches a minimum of $\kindex{CoT}{free,min}=\SI{29.8}{\joule/\meter}$ for free swimming and $\kindex{CoT}{teth,min}=\SI{23.8}{\joule/\meter}$ for tethered swimming.
Beyond this point, the cost of transport increases again as the specific tail amplitude grows.
The minimum values of the cost of transport occur approximately at the same values of the specific tail amplitude, but the overall values differ. 
The cost of transport for free swimming rapidly increases for tail amplitudes away from the optimum and reach values up to $\kindex{CoT}{free,max}=\SI{80}{\joule/\meter}$.
The cost of transport for tethered swimming is confined to a narrower range with $\kindex{CoT}{teth,max}=\SI{33}{\joule/\meter}$.
Overall, the cost of transport is higher for free than for tethered swimming (\cref{fig:freevstetheredkin}c).

The slip ratio $\kindex{U/V}{wave}$ provides a complementary measure of propulsive performance \cite{Stin2024}.  
For free swimming, the minimum cost of transport occurs near $\kindex{U/V}{wave}\approx0.55$, whereas the maximum observed value of approximately $0.70$ corresponds to suboptimal energetic performance (\supfig{7}).

The analysis of the kinematics explains the similarities in stride length between tethered and free swimming. 
The similarity is attributed to the similar tail motion that governs the stride length behaviour. 
As we will see next, the discrepancies in the cost of transport are deemed related to the fixation of the head and the suppression of roll motion and lateral head motion in the tethered swimming configuration.

\subsubsection{Mechanical stabilisation by tethering}

\begin{figure}[!t]
\centering\includegraphics{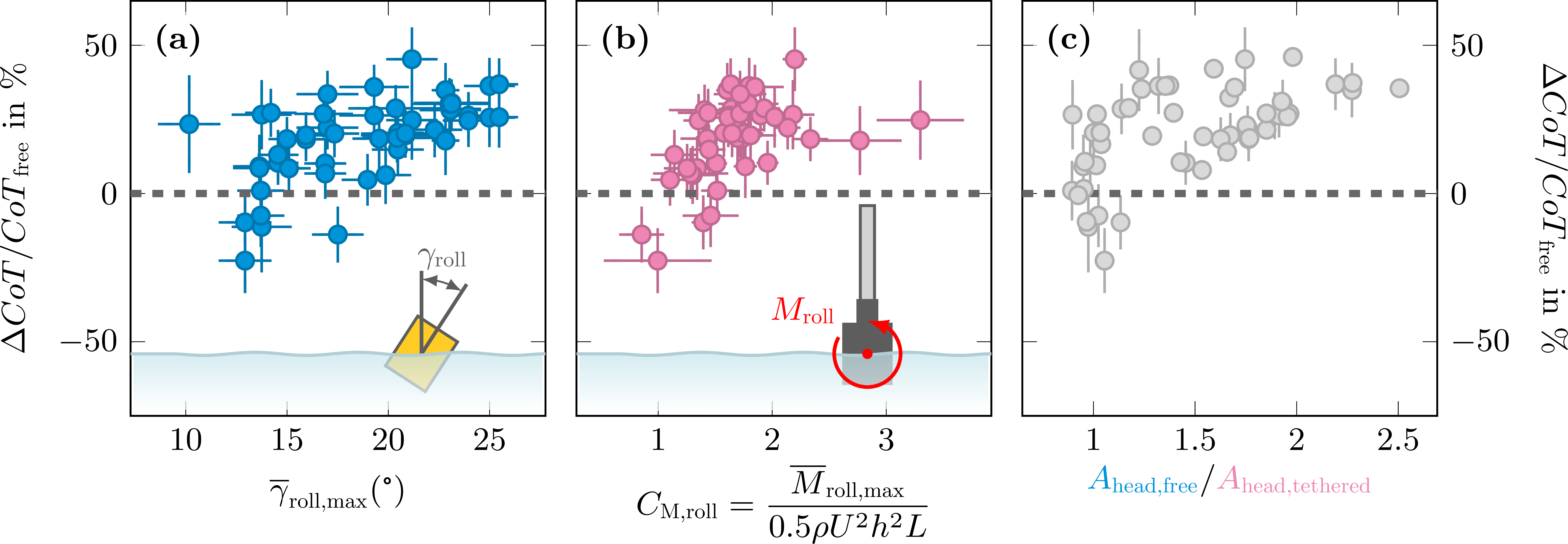}
\caption{
Tethered efficiency gain ($\Delta CoT/\kindex{CoT}{free} = (\kindex{CoT}{free}-\kindex{CoT}{tethered})/\kindex{CoT}{free}$) as a function of the mechanical stabilisation imposed by tethering.
Maximum roll is obtained (a) directly from measuring the roll angle \kindex{\gamma}{roll} during free swimming and (b) indirectly through the rolling moment \kindex{M}{roll} measured during tethered experiments and presented in the form of the maximum rolling moment coefficient \kindex{C}{M,roll}. 
%Colours represent the free versus tethered swimming calculation of maximum roll.
(c) Tethered efficiency gain as a function of the relative lateral head motion suppression expressed as the head amplitude ratio: $\kindex{A}{head,tethered}/\kindex{A}{head,free}$.
}
\label{fig:freevstetheredroll}.
\end{figure}

Tethering modifies several motions that are otherwise permitted during free swimming.
The comparison of the kinematics shows that the head amplitude is reduced in the tethered configuration. 
The relative suppression of the lateral head motion is quantified by the head amplitude ratio: $\kindex{A}{head,tethered}/\kindex{A}{head,free}$.

The roll motion around the main swimming axis is also observed in several free-swimming trials. %(see \supvid{tba}).% @AA: add video name
Body roll is a natural component of swimming kinematics across species \cite{Shepard2010}, where it can arise from body–fluid interactions and is often actively regulated.
In undulatory swimmers, roll has been linked to stability and control: semi-aquatic snakes adjust body immersion to limit rolling near the surface \cite{Herault2020}, while fish regulate body orientation through morphological features such as fins and body shape \cite{Eidietis2002,Webb2002}.
% In our robotic system, roll introduces an additional degree of freedom that does not contribute to forward propulsion and dissipates energy.

In our robotic configurations, the roll and lateral head amplitude are suppressed in the tethered configuration, which are more efficient in general. 
We conducted a quantitative analysis to get more insights into the role of mechanical stabilisation on the observed efficiency difference between free and tethered swimming.
For free swimming, the onboard accelerometer was used to estimate the roll angle evolution \kindex{\gamma}{roll}.
The roll amplitude \kindex{\overline{\gamma}}{roll,max} is defined as the mean over five undulations of the amplitude of $\gamma$.
For tethered swimming, roll was estimated indirectly from the roll moment \kindex{M}{roll} measured by the load cell during tethered swimming.
The roll moment amplitude \kindex{\overline{M}}{roll,max} is defined as the mean of the maximum value of the amplitude of \kindex{M}{roll} for five successive undulations.
The roll moment coefficient \kindex{C}{M,roll,max} is defined as the roll moment amplitude normalised by $0.5\rho U^2h^2L$.
The tethered efficiency gain between free and tethered swimming was defined as $\Delta CoT/\kindex{CoT}{free} = (\kindex{CoT}{free}-\kindex{CoT}{teth})/\kindex{CoT}{free}$ for kinematics with the same input parameters for a quantitative comparison.

% Description of the results
The tethered efficiency gain is shown in \cref{fig:freevstetheredroll} as a function of the free-swimming roll-angle amplitude, the tethered roll-moment coefficient, and the relative head amplitude suppression.
The free-swimming roll angles span $\kindex{\overline{\gamma}}{roll,max} =\SIrange{10}{25}{\degree}$, with a mean of $\ang{17}$ (\cref{fig:freevstetheredroll}a).
The tethered efficiency gain spans from $\SIrange{-22}{46}{\%}$, with a median value of $20\%$ and only five negative cases.
Overall, tethering improves the efficiency of the robot.
An increase in roll results in an increase in the tethered efficiency gain, for both direct and indirect measurements. 
The tethered efficiency gain also increases with the head amplitude ratio $\kindex{A}{head,tethered}/\kindex{A}{head,free}$ (\cref{fig:freevstetheredroll}c), following a trend similar to that observed for roll.
By contrast, the lateral centre-of-mass sway measured during free swimming shows no clear relationship with the tethered efficiency gain (\supfig{9}), despite being mechanically suppressed in the tethered
configuration \cite{Xiong2014}.
When tethering suppresses high roll motions and reduces the anterior lateral motion, the efficiency of locomotion improves because less energy is converted into body rotation and the drag penalty associated with anterior undulation is suppressed.

The tethered efficiency gain increases with both free-swimming roll and relative head amplitude suppression.
For high roll and large head amplitude reduction kinematics, the tethering stabilises the robot, reduces energy expenditure, and likely diminishes wave creation that could contribute to enhanced drag. 
For kinematics with low roll and low head amplitude reduction, the energetic benefit of tethering is reduced and can become negative.
The present measurements do not allow us to identify the mechanism responsible for the cases where tethering has a negative effect.
Because roll and lateral head motion are modified simultaneously by the shaft mounting, their individual energetic contributions cannot be isolated using the present experiments.

Overall, kinematics prone to roll and large head amplitudes benefit from tethering, whereas inherently roll-stable and head-stable kinematics can experience a loss of efficiency.
Recent robotic developments aim to address this limitation by reintroducing roll dynamics: some systems allow partial three-dimensional motion using flexible tethers \cite{White2021,Zhu2019}, while others incorporate an additional degree of freedom around the roll axis to actively control body rotation \cite{Xie2023}.
These approaches mirror biological strategies, where swimmers regulate roll through active control or morphological adaptations.

\subsection{Depth variation: from surface to fully submerged swimming}

\subsubsection{Performance overview}
\begin{figure}[t]
\centering\includegraphics{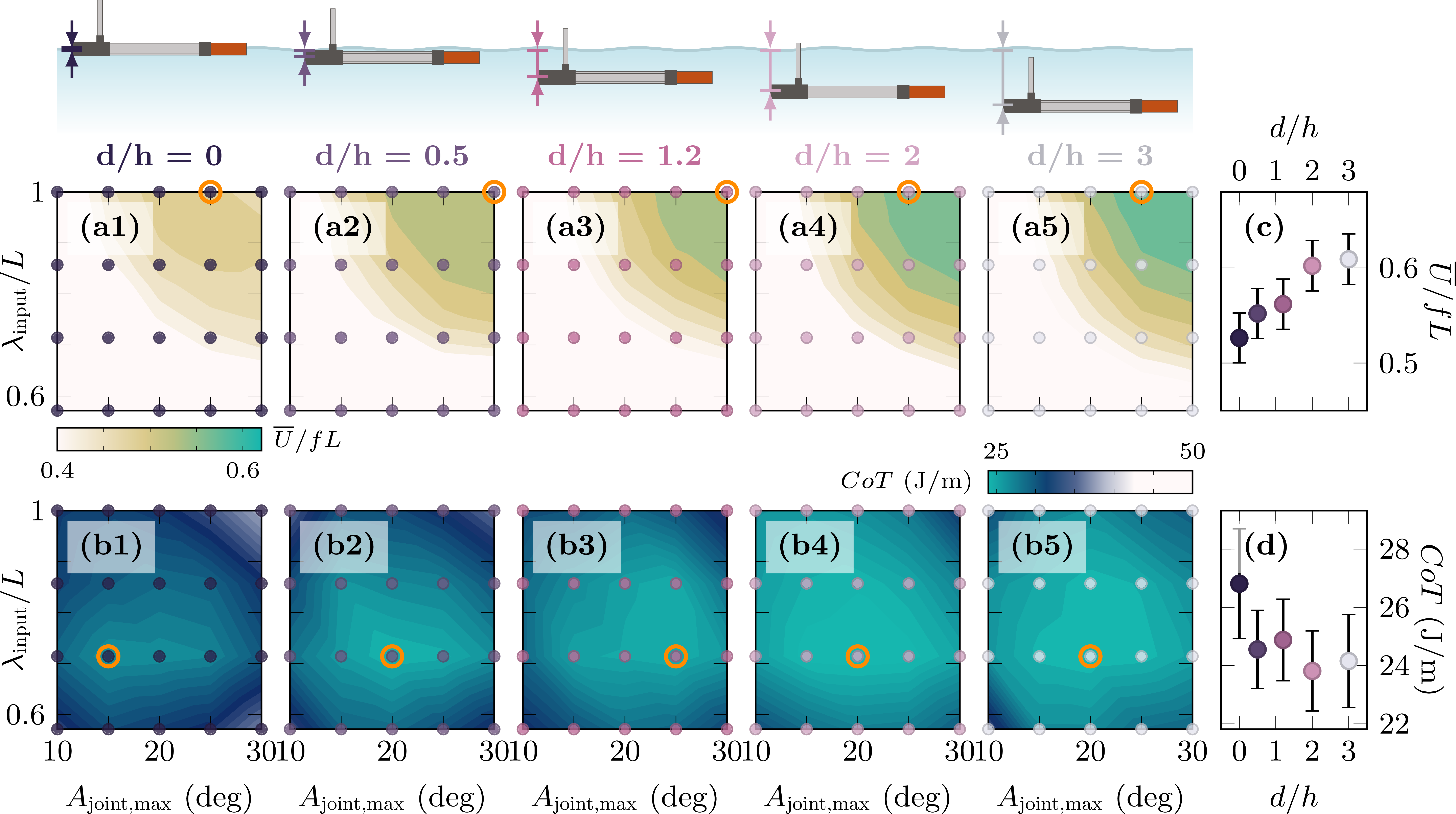}
\caption{
Tethered swimming performance maps of normalised stride length (a) and cost of transport (b) for increasing submergence depths (left to right) as a function of the input kinematic parameters ($\kindex{\lambda}{input}/L$ and $\kindex{A}{joint, max}$). Orange circles indicate the best performance points for every map. The corresponding optimal values of stride length (c) and cost of transport (d) are shown as functions of submergence depth.
}
\label{fig:surfacevsdeepmaps}
\end{figure}

Submergence depth is known to influence swimming performance through interactions with the free surface.  
Experimental studies on live fish have shown that swimming near the surface can reduce performance due to energy dissipation through wave generation \cite{Webb1991}.  
Similar behaviour has been reported for bio-inspired systems: flexible oscillating plates exhibit increased thrust and efficiency with increasing submergence depth, as surface-induced energy losses diminish \cite{Barannyk2012}.  
Rigid oscillating foils show comparable trends, although flow visualisations indicate that the free surface can also act as a symmetry plane, partially recovering thrust under specific conditions \cite{Huera-Huarte2023}.  
In addition, unsteady interactions between the body motion and surface waves can lead to both performance losses and gains depending on operating conditions, particularly near resonance \cite{Dode2022}.  

The effect of submergence depth on tethered swimming performance is examined here.
Performance is expressed again by the normalised stride length ($\overline{U}/fL$) and the cost of transport $CoT$.
\Cref{fig:surfacevsdeepmaps} summarises the performance maps as a function of the input kinematic parameters ($\kindex{\lambda}{input}/L$ and $\kindex{A}{joint,max}$) for swimming at different submergence depths varying from the surface to a depth of three times the height of the robot. 

The stride length distributions are shown in \cref{fig:surfacevsdeepmaps}a for all the tested depths and input parameter combinations.
For low input parameters $\kindex{A}{joint,max}<\ang{20}$ and $\kindex{\lambda}{input}/L<0.74$, stride length increased with both $\kindex{A}{joint,max}$ and $\kindex{\lambda}{input}/L$. 
An optimum region is reached for higher values of input parameters.
The topology of the resulting contour plots is similar for all depths examined here.
The maximum normalized stride length increases with depth, from $\kindex{U/fL}{max} (d/h=0)=0.54$ at the surface to successively: $\kindex{U/fL}{max} (d/h=0.5)=0.56$, $\kindex{U/fL}{max} (d/h=1.2)=0.57$, $\kindex{U/fL}{max} (d/h=2)=0.60$, and $\kindex{U/fL}{max} (d/h=3)=0.61$.
Beyond $d/h=2$, the maximum value of the stride length converges, indicating that submerging the robot further does not alter its performance.
Overall, the maximum stride length increases with increasing depth, even though the topology of performance maps is similar. 

The cost of transport distributions are shown in \cref{fig:surfacevsdeepmaps}b for the tested depths and input parameters.
For all depths, the cost of transport exhibits a minimum value within a specific range of input parameters.
At the surface, the optimum region appears as a narrow band centred around $\kindex{\lambda}{input}\approx \numrange{0.72}{0.75}$ and $\kindex{\lambda}{input}\approx \SIrange{14}{25}{\degree}$ (\cref{fig:surfacevsdeepmaps}b1).
The optimum region gradually increases with increasing depth, reaching a range of $\kindex{\lambda}{input}\approx \numrange{0.72}{0.88}$ and $\kindex{\lambda}{input}\approx \SIrange{12}{28}{\degree}$ at $d/h=3$ (\cref{fig:surfacevsdeepmaps}b4).
Input parameters outside these ranges increase the cost of transport. 
The minimum $CoT$ value decreases for increasing depth from $\kindex{CoT}{min} (d/h=0)=\SI{26.8}{\joule/\meter}$ at the surface and successively: $\kindex{CoT}{min} (d/h=0.5)=24.6$, $\kindex{CoT}{min} (d/h=1.2)=24.8$, $\kindex{CoT}{min} (d/h=2)=23.8$, $\kindex{CoT}{min} (d/h=3)=24.1$.
As for the stride length, the minimum value of the $CoT$ reaches a converged value for the last two depths tested, indicating that further increase in depth does not affect the performance.
Overall, as depth increases, the minimum cost of transport values decrease, and the optimum region of minimum cost of transport spread to a wider range of input values. 

\subsubsection{Kinematic insights}
\begin{figure}[t]
\centering\includegraphics{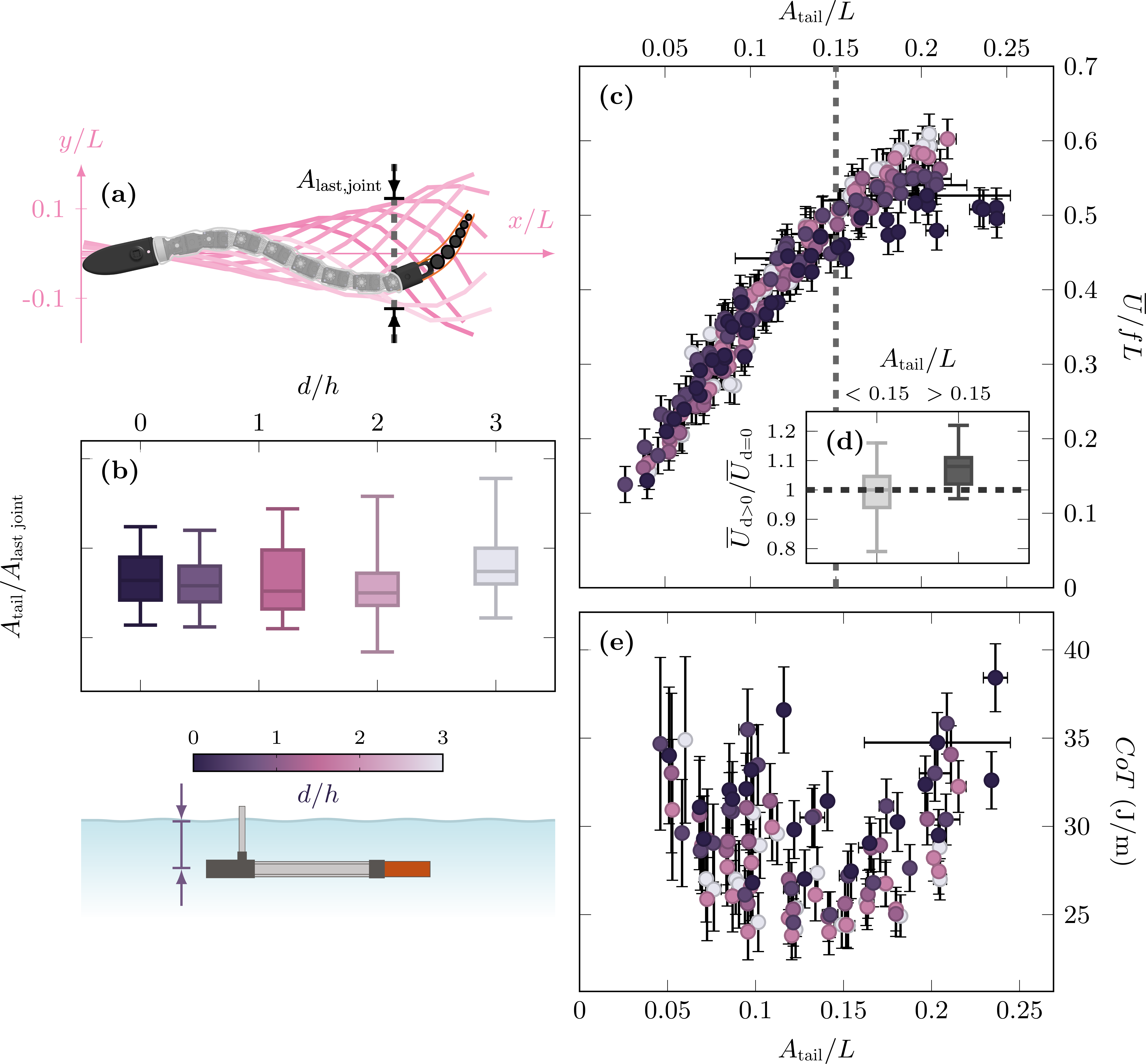}
\caption{Kinematic analysis for tethered swimming at various depths.
(a) Example indicating the extraction of the last joint amplitude \kindex{A}{last joint}.
(b) Statistics of the ratio of the tail amplitude to the amplitude of the last joint: $\kindex{A}{tail}/\kindex{A}{last joint}$ as a function of depth.
(c) Normalised stride length as a function of the tail amplitude.
(d) Statistics of the ratio of speed for submerged swimming \kindex{\overline{U}}{d\textgreater0} to speed for surface swimming \kindex{\overline{U}}{d=0} for low tail amplitudes ($\kindex{{A}}{tail}/L<0.15$) and high tail amplitudes ($\kindex{{A}}{tail}/L>0.15$).
%(e) Normalised stride length as a function of the specific tail amplitude across depths.
(e) Cost of transport as a function of the specific tail amplitude across depths.
}
\label{fig:surfacevsdeepkin}
\end{figure}

% stride length
The effect of the resulting kinematics on the performance across depths is summarised in \cref{fig:surfacevsdeepkin}.
To compare swimming kinematics across depths, we take the amplitude of the last joint of the robot as reference.
The amplitude of the last joint is calculated similarly to the tail amplitude as shown in \cref{fig:surfacevsdeepkin}a. 
We define the ratio of the tail amplitude to the amplitude of the last joint as $\kindex{A}{tail}/\kindex{A}{last joint}$.
This ratio compares the passive deflection of the tail to the last part of the body directly controlled by the input parameters. 

Statistical distributions of the amplitude ratio $\kindex{A}{tail}/\kindex{A}{last joint}$ are shown in \cref{fig:surfacevsdeepkin}b as a function of depth.
The median value of the amplitude ratios varies from $1.82$ at the surface and subsequently $1.79$ at $d/h=0.5$, $1.76$ at $d/h=1.2$, $1.75$ at $d/h=2$, and $1.87$ at $d/h=3$.
The median value of the amplitude ratio does not present any increasing or decreasing trend as a function of depth.
A supplementary analysis of the recorded kinematics shows no significant depth-dependent change for the last joint amplitude (\kindex{A}{last joint}) or for the angle amplitude tracking across all eight joint angles (\supfig{8}).
In conclusion, the variation of depth has no significant effect on the performed kinematics.

The stride length is shown in \cref{fig:surfacevsdeepkin}c as a function of the tail amplitude for all depths tested. 
For small tail amplitudes ($\kindex{{A}}{tail}/L<0.15$), an increase in the tail amplitude leads to an increase in stride length with similar values of \kindex{A}{tail} resulting in similar $\overline{U}/fL$ across depths. 
At larger tail amplitudes ($\kindex{{A}}{tail}/L>0.15$), the stride length continues to increase with increasing $\kindex{A}{tail}$ when fully submerged.
When swimming closer to the surface, the increase in stride length with tail amplitude is reduced compared to the fully submerged swimming for the same tail amplitude. 
This results in a plateau in stride length for the surface swimming ($d/h=0$), where an increase in amplitude does not further increase the stride length.
To illustrate the difference in speed for amplitudes greater than $0.15$, the ratio of the speed for submerged swimming \kindex{\overline{U}}{d\textgreater0} and the speed for surface swimming \kindex{\overline{U}}{d=0} is calculated for experiments with the same kinematic inputs.
The distribution of the speed ratio $\kindex{\overline{U}}{d\textgreater0}/\kindex{\overline{U}}{d=0}$ is presented in \cref{fig:surfacevsdeepkin}d for the two conditions: $\kindex{{A}}{tail}/L<0.15$ and $\kindex{{A}}{tail}/L>0.15$.
The speed ratio for $\kindex{{A}}{tail}/L<0.15$ results in a median value of unity with an interquartile range of \numrange{0.94}{1.05}.
The median value of unity illustrates that depth does not affect the swimming speed for low values of $\kindex{{A}}{tail}$.
The speed ratio for $\kindex{{A}}{tail}/L>0.15$ results in a median value of $1.08$ with an interquartile range of \numrange{1.02}{1.11}.
The median value of $1.08$ illustrates that speed decreased near the surface for high values of $\kindex{{A}}{tail}$.
Overall, stride length increases with tail amplitude across depths, but the near-surface condition limits performance at large amplitudes.

Cost of transport as a function of specific tail amplitude is presented in \cref{fig:surfacevsdeepkin}e.
Cost of transport decreases with increasing amplitude until $\kindex{A}{tail}/L\approx\numrange{0.10}{0.15}$, where it reaches its minimum for all depths.
These optimal amplitudes occur below the values corresponding to maximum stride length.
Further increases in amplitude increase the cost of transport again, indicating a trade-off between thrust generation and energetic efficiency.

\section{Concluding remarks}

This study characterised the performance of the undulatory robot 1-guilla during free and tethered swimming at the surface and during tethered swimming at different submergence depths, covering conditions ranging from free surface swimming to tethered fully submerged swimming.
Performance and kinematic analyses revealed how mechanical constraints and surface proximity affect swimming efficiency, addressing two main questions: 
1)~What are the performance differences between free-swimming and tethered configurations?
2)~How do surface versus fully submerged swimming conditions affect the experimental outcomes?

\subsection{The effect of tethering}

Surface swimming performance was characterised for the robot swimming freely in the pool and tethered in the water channel.
Free swimming reached comparable swimming speeds to tethered swimming.
The stride length increased with an increase in specific tail amplitude ($\kindex{A}{tail}/(\lambda/4)$) until unity. 
A further increase of $\kindex{A}{tail}/(\lambda/4)$ decreased the stride length. 
The tethering did not affect the behaviour of the tail, resulting in similar tail amplitudes for the same input parameter values.
As the tail governs the swimming performance in terms of the stride length \cite{Anastasiadis2023}, and as long as tethering does not alter the tail motion, swimming speed remains unaffected.

Efficiency in terms of cost of transport was quantified for surface swimming and linked to the mechanical stabilisation imposed by tethering.
The tethering of the robot leads to an average decrease of $20.5\%$ in the energy consumption for the same kinematic inputs across our data. 
The decrease in energy consumption (or increase in efficiency) is higher for kinematics exhibiting larger roll and lateral head amplitudes during free-swimming (\cref{fig:freevstetheredroll}). 
Maintaining roll and whole-body stability appears to be essential for efficient propulsion near the free surface, pointing to the potential need for active control strategies \cite{Herault2020} or morphological adaptations such as enlarged fins or body stiffness gradients \cite{Eidietis2002}. 
Stabilisation of the head also appears essential for efficient propulsion. 
Mechanical stabilisation likely suppresses the drag penalty associated with anterior body undulation, allowing the tethered swimmer to match free-swimming speeds at lower costs of transport.
Because roll and lateral head motion are constrained simultaneously, their individual energetic contributions cannot be separated in the present configuration.
Biological swimmers may employ specialised mechanisms to counteract this effect and maintain their efficiency. 

\subsection{The effect of submergence depth}
The performance of the robot was evaluated for tethered swimming at varying depths, from the surface to fully submerged.
At small tail amplitudes, the stride length was similar across depths. 
At large amplitudes, swimming close to the surface produced a plateau of lower stride length compared to deep swimming. 
For the most submerged condition tested, the stride length was up to $13\%$ higher than at the surface for the range of kinematics tested.
The same performance improvement of $12\%$ is observed for the optimum values of the resulting cost of transport when transitioning from the surface to deeply submerged.
Not only does the optimum value of cost of transport improve, but the performance maps also show a larger area in which cost-efficient kinematics exist (\cref{fig:surfacevsdeepmaps}).
Resulting kinematics were similar for similar input parameters from deeply submerged up to the surface.
Surface proximity had a clear influence on swimming performance despite unchanged kinematics.
Since the kinematic parameters remained similar, the loss of speed and efficiency is attributed to surface wave generation.
Wave generation is often linked to performance losses associated with this type of unsteady motion.

Both forward translation and lateral body motion may contribute to free-surface wave generation.
The maximum translational Froude number based on the body length, defined as:
\begin{equation}
\kindex{Fr}{U,L}=\overline{U}/\sqrt{gL}\quad,
\end{equation}
remains below $\kindex{Fr}{U,L}=0.18$.
The Froude number based on the maximum lateral tail velocity
\begin{equation}
\kindex{Fr}{U,lat}=2\pi\kindex{A}{tail}/\sqrt{gL}\quad,
\end{equation}
reaches values up to \num{0.45}.
The corresponding depth-based Froude values 
\begin{align}
\kindex{Fr}{U,d}&=\overline{U}/\sqrt{g(d+h/2)}\\
\kindex{Fr}{{lat},d}&=2\pi f\kindex{A}{tail}/\sqrt{g(d+h/2)}\quad,
\end{align}
reach maximum values of $\kindex{Fr}{U,d}=0.8$ and $\kindex{Fr}{{lat},d}=2$.
The larger Froude numbers associated with the lateral motion, together with the stronger surface-induced performance loss at large tail amplitudes, suggests that body and tail oscillations may contribute substantially to wave generation.
Forward translation may also contribute to wave generation near the surface, but direct measurements of the wave field would be required to separate changes in the wave drag from changes in the propulsive thrust~\cite{Huera-Huarte2023,Polly2025,Rocuzzo2026}.

Unfortunately, measurements for a free-swimming submerged robot were not feasible in this study, which limits our ability to completely decouple the effects due to mechanical constraints from depth-dependent hydrodynamic effects.
The present experimental design quantifies the effect of tethering only at the surface and the effect of submergence depth only in the tethered configuration.
The $20.5\%$ reduction in energetic cost associated with tethering cannot be assumed to persist at depth.
A complete comparison would require a swimming pool with sufficient depth and a freely swimming robot with neutral buoyancy and attitude control to maintain a prescribed submerged trajectory.

\subsection{Guidelines for experimentalists}

The present results provide practical guidelines for interpreting and comparing undulatory swimming experiments across free, tethered, and surface conditions.  

First, trends in performance with respect to input kinematics are preserved across configurations.  
Variations in parameters such as joint amplitude or wavelength lead to consistent changes in stride length and cost of transport in both free and tethered swimming.  
For example, maximum stride length was consistently achieved at a specific tail amplitude of $\kindex{A}{tail}/(\lambda/4)=1$ across all configurations.  
As a result, tethered experiments can reliably capture relative performance trends, provided that the resulting kinematics—particularly tail motion—are not significantly altered.  

Second, absolute performance values must be interpreted with caution.
At the surface, tethering can artificially improve efficiency by suppressing roll and lateral head motion, leading to a reduction in cost of transport of approximately $20\%$ for identical kinematic inputs.
Under tethered conditions, proximity to the surface also reduces performance: maximum stride length and optimal efficiency are reduced by approximately $12$--$13\%$ compared to deeply submerged conditions.
The absence of results from a fully submerged, freely swimming robot does not allow us to completely decouple the effects due to mechanical constraints from effects due to a varying submergence depth. 
Comparisons across configurations are meaningful only when the underlying kinematics remain comparable, and the effects of constraints are properly accounted for.  

Third, the design of experimental setups plays a critical role in preserving physical realism.  
Rigid tethering systems that constrain rotational degrees of freedom can alter body kinematics and bias performance measurements.
Constraining rotation about the body axis suppresses roll, while fixing the lateral position near the head reduces anterior lateral motion.
Both effects can alter the measured energetic performance.
Rotation in the horizontal plane (yaw) is typically preserved in shaft-based setups; suppressing this degree of freedom would be expected to induce significant changes in body kinematics and propulsion.
Constraints on other rotational degrees of freedom, such as pitch, are less pronounced in the present configuration but may still influence body orientation and local flow interactions. 
Allowing additional degrees of freedom or minimising mechanical constraints improves the agreement with free-swimming behaviour and reduces systematic bias in efficiency estimates.  
The sensitivity of energetic performance to roll and lateral head motion suppression highlights the importance of three-dimensional body control in maintaining efficient propulsion.  
Overall, mechanical stabilisation by tethering and submergence depth emerge as key factors governing undulatory swimming performance, with effects on energy efficiency and speed on the order of $10$--$20\%$.  

Finally, these findings offer insight into biological swimming strategies.  
The reduced performance observed near the surface supports the tendency of many fish to avoid surface locomotion \cite{Videler1993}.
The observed tail-amplitude optima can be compared with results from selected kinematics observed in biological anguilliform swimmers.
The maximum stride length for our robot occurs at a specific tail amplitude of $\kindex{A}{tail}/(\lambda/4)=1$, whereas minimum cost of transport occurs near $\kindex{A}{tail}/(\lambda/4)\approx0.65$.
Free-swimming eels select a specific tail amplitude of approximately $0.55$ \cite{Tytell2004a,Anastasiadis2023}, which lies closer to the energetic optimum than to the maximum-speed condition.
In our experiments, we measure tail amplitudes $\kindex{A}{tail}/L\approx0.1$, which would fall in the range associated with minimum cost of transport identified by \cite{Sanchez-Rodriguez2021}.
These comparisons suggest that the kinematics observed during steady anguilliform swimming might be more targeting efficiency than maximum speed.
Biological swimmers can actively modulate body stiffness, whereas the present robot has a fixed mechanical stiffness, which limits a direct quantitative comparison.

\ack{
Authors thank Pauline Nicolas, Ga\'etan Raynaud, Sahar Rezapour and Mrudhula Baskaran for the valuable help during data gathering; 
Alessandro Crespi, Fran\c{c}ois Longchamp for the technical support; 
Ga\'etan Raynaud for the photographic assistance, discussions, and feedback;
Xiangxiao Liu for the discussions about the waterproofing of the tethered version of the robot;
The personnel of the EPFL mechanical workshop for technical support. 
}

% \funding{
% }
% This section is a list of funder names and grant numbers

\roles{
\textbf{Alexandros Anastasiadis}: conceptualisation, formal analysis, investigation, software, writing - original draft, visualisation; 
\textbf{Auke J. Ijspeert}: conceptualisation, writing - review \& editing, supervision, funding acquisition; 
\textbf{Karen Mulleners}: conceptualisation, writing - review \& editing, supervision, funding acquisition;
}
% List author names and the contributions made to the article, using terms from the NISO Contributor Roles Taxonomy (CRediT) https://credit.niso.org

\data{All data available in this paper will be made available on Zenodo prior to publication.}
% For more information on IOP Publishing's research data policy see: https://publishingsupport.iopscience.iop.org/questions/research-data/

\suppdata{See Supplementary Information pdf and Supplementary Videos}

%%%%%%%%%% bibliography %%%%%%%%%%%%%%

\bibliographystyle{plain}  % or another style like ieeetr, apalike, unsrt, etc.
\bibliography{references}

@article{mulleners.2024,
  title = {Self-Exploring Automated Experiments for Discovery, Optimization, and Control of Unsteady Vortex-Dominated Flow Phenomena},
  author = {Mulleners, Karen},
  year = {2024},
  journal = {Physical Review Fluids},
  shortjournal = {Phys. Rev. Fluids},
  volume = {9},
  number = {12},
  pages = {124701},
  publisher = {American Physical Society},
  doi = {10.1103/PhysRevFluids.9.124701},
}

@article{Xiong2014,
	author = {Xiong, Grace and Lauder, George V},
	doi = {10.1016/j.zool.2014.03.002},
	issn = {1873-2720 (Electronic)},
	journal = {Zoology (Jena, Germany)},
	language = {eng},
	month = {aug},
	number = {4},
	pages = {269--281},
	pmid = {24925455},
	title = {{Center of mass motion in swimming fish: effects of speed and locomotor mode  during undulatory propulsion.}},
	volume = {117},
	year = {2014}
}

@article{Polly2025,
	author = {Polly, Gatien and M{\'{e}}rigaud, Alexis and Thiria, Benjamin and Godoy-Diana, Ramiro},
	doi = {DOI: 10.1017/jfm.2025.90},
	edition = {2025/03/14},
	issn = {0022-1120},
	journal = {Journal of Fluid Mechanics},
	pages = {R4},
	publisher = {Cambridge University Press},
	title = {{Experiments on water-wave interactions with a horizontal submerged elastic plate}},
	url = {https://www.cambridge.org/core/product/211D6233CD7EB3F291CDA5A798539B4B},
	volume = {1007},
	year = {2025}
}

@article{Obayashi2026,
	author = {Obayashi, Nana and Anastasiadis, Alexandros and Gumowski, Jessica and Junge, Kai and Walker, Kyle L and Mulleners, Karen and Hughes, Josie},
	doi = {10.1038/s44182-026-00105-z},
	issn = {2731-4278},
	journal = {npj Robotics},
	title = {{ScaFi: length-scalable, compliant, parametric robotic fish design for operation in multiple environmental niches}},
	url = {https://doi.org/10.1038/s44182-026-00105-z},
	year = {2026}
}

@article{Rocuzzo2026,
	title = {Transitions in unsteady capillary-gravity wakes of surface swimmers},
	author = {Roccuzzo, Max and Herault, Johann},
	journal = {Phys. Rev. Fluids},
	volume = {11},
	issue = {4},
	pages = {044805},
	numpages = {27},
	year = {2026},
	month = {Apr},
	publisher = {American Physical Society},
	doi = {10.1103/srzb-5ksp},
	url = {https://link.aps.org/doi/10.1103/srzb-5ksp}
}

@book{Videler1993,
author = {Videler, John J.},
booktitle = {Fish Swimming},
publisher = {Springer Dordrecht},
doi = {10.1007/978-94-011-1580-3},
isbn = {9789401046879},
title = {{Fish Swimming}},
year = {1993}
}

@book{Conte2017,
	address = {Philadelphia, Pennsylvania},
	author = {Conte, Samuel Daniel and {De Boor}, Carl},
	doi = {10.1137/1.9781611975208},
	edition = {3d ed.},
	isbn = {1-61197-520-4},
	publisher = {Society for Industrial and Applied Mathematics SIAM},
	title = {{Elementary numerical analysis : an algorithmic approach }},
	year = {2017},
	page = {76}
}

@article{Huera-Huarte2023,
author = {Huera-Huarte, Francisco J.},
doi = {10.1016/j.oceaneng.2023.113663},
issn = {00298018},
journal = {Ocean Engineering},
number = {January},
publisher = {Elsevier Ltd},
title = {{Pitching foil propulsion performance decays near the free surface}},
volume = {272},
year = {2023}
}

@article{Dode2022,
author = {Dode, A. and Carmigniani, R. and Cohen, C. and Clanet, C. and Bocquet, L.},
doi = {10.1017/jfm.2022.592},
issn = {14697645},
journal = {Journal of Fluid Mechanics},
pages = {0--27},
title = {{Wave drag during an unsteady motion}},
volume = {951},
year = {2022}
}

@article{Barannyk2012,
author = {Barannyk, Oleksandr and Buckham, Bradley J. and Oshkai, Peter},
doi = {10.1016/j.jfluidstructs.2011.10.005},
issn = {08899746},
journal = {Journal of Fluids and Structures},
pages = {152--166},
publisher = {Elsevier},
title = {{On performance of an oscillating plate underwater propulsion system with variable chordwise flexibility at different depths of submergence}},
url = {http://dx.doi.org/10.1016/j.jfluidstructs.2011.10.005},
volume = {28},
year = {2012}
}

@article{Webb1991,
author = {Webb, P. W. and Sims, D. and Schultz, W. W.},
doi = {10.1242/jeb.155.1.219},
issn = {00220949},
journal = {Journal of Experimental Biology},
pages = {219--226},
title = {{The effects of an air/water surface on the fast-start performance of rainbow trout (Oncorhynchus mykiss)}},
volume = {155},
year = {1991}
}

@article{Triantafyllou1995,
author = {Triantafyllou, Michael S. Triantafyllou and George S.},
journal = {Scientific American},
number = {3},
pages = {64--70},
title = {{An Effcient Swimming Machine}},
volume = {272},
year = {1995}
}

@misc{Hultmark2010,
author = {Hultmark, Marcus and Leftwich, Megan and Smits, Alexander J.},
booktitle = {Animal Locomotion},
doi = {10.1007/978-3-642-11633-9_5},
isbn = {9783642116322},
pages = {45--52},
title = {{Flowfield measurements in the wake of a robotic lamprey}},
year = {2010}
}

@article{Curet2011,
author = {Curet, Oscar M. and Patankar, Neelesh A. and Lauder, George V. and MacIver, Malcolm A.},
doi = {10.1088/1748-3182/6/2/026004},
issn = {17483182},
journal = {Bioinspiration and Biomimetics},
number = {2},
pmid = {21474864},
title = {{Mechanical properties of a bio-inspired robotic knifefish with an undulatory propulsor}},
volume = {6},
year = {2011}
}

@article{Salumae2013,
author = {Salum{\"{a}}e, Taavi and Kruusmaa, Maarja},
doi = {10.1098/rspa.2012.0671},
issn = {14712946},
journal = {Proceedings of the Royal Society A: Mathematical, Physical and Engineering Sciences},
number = {2153},
title = {{Flow-relative control of an underwater robot}},
volume = {469},
year = {2013}
}

@article{Chambers2014,
author = {Chambers, L. D. and Akanyeti, O. and Venturelli, R. and Jezǒv, J. and Brown, J. and Kruusmaa, M. and Fiorini, P. and Megill, W. M.},
doi = {10.1098/rsif.2014.0467},
issn = {17425662},
journal = {Journal of the Royal Society Interface},
number = {99},
pmid = {25079867},
title = {{A fish perspective: Detecting flow features while moving using an artificial lateral line in steady and unsteady flow}},
volume = {11},
year = {2014}
}

@article{Jusufi2017,
author = {Jusufi, Ardian and Vogt, Daniel M. and Wood, Robert J. and Lauder, George V.},
doi = {10.1089/soro.2016.0053},
issn = {21695180},
journal = {Soft robotics},
number = {3},
pages = {202--210},
pmid = {29182079},
title = {{Undulatory Swimming Performance and Body Stiffness Modulation in a Soft Robotic Fish-Inspired Physical Model}},
volume = {4},
year = {2017}
}

@article{Gibouin2018,
author = {Gibouin, Florence and Raufaste, Christophe and Bouret, Yann and Argentina, M{\'{e}}d{\'{e}}ric},
doi = {10.1063/1.5043137},
issn = {10897666},
journal = {Physics of Fluids},
number = {9},
title = {{Study of the thrust-drag balance with a swimming robotic fish}},
volume = {30},
year = {2018}
}

@article{Zheng2018,
author = {Zheng, Xingwen and Wang, Chen and Fan, Ruifeng and Xie, Guangming},
doi = {10.1088/1748-3190/aa8f2e},
issn = {17483190},
journal = {Bioinspiration and Biomimetics},
number = {1},
pmid = {28949301},
publisher = {IOP Publishing},
title = {{Artificial lateral line based local sensing between two adjacent robotic fish}},
volume = {13},
year = {2018}
}

@article{Zhong2021,
author = {Zhong, Q. and Zhu, J. and Fish, F. E. and Kerr, S. J. and Downs, A. M. and Bart-Smith, H. and Quinn, D. B.},
doi = {10.1126/scirobotics.abe4088},
issn = {24709476},
journal = {Science Robotics},
number = {57},
pmid = {34380755},
title = {{Tunable stiffness enables fast and efficient swimming in fish-like robots}},
volume = {6},
year = {2021}
}

@article{Sanchez-Rodriguez2021,
author = {S{\'{a}}nchez-Rodr{\'{i}}guez, J. and Celestini, F. and Raufaste, C. and Argentina, M.},
doi = {10.1103/PhysRevLett.126.234501},
issn = {10797114},
journal = {Physical Review Letters},
number = {23},
pages = {234501},
publisher = {American Physical Society},
title = {{Proprioceptive Mechanism for Bioinspired Fish Swimming}},
url = {https://doi.org/10.1103/PhysRevLett.126.234501},
volume = {126},
year = {2021}
}

@article{Stin2024,
	author = {Stin, Vincent and Godoy-diana, Ramiro and Bonnet, Xavier and Herrel, Anthony},
	doi = {10.1111/brv.13116},
	journal = {Biological Reviews},
	pages = {2190--2210},
	title = {{Form and function of anguilliform swimming}},
	volume = {6},
	year = {2024}
}

@article{Thandiackal2021,
author = {Thandiackal, Robin and Melo, Kamilo and Paez, Laura and Herault, Johann and Kano, Takeshi and Akiyama, Kyoichi and Boyer, Fr{\'{e}}d{\'{e}}ric and Ryczko, Dimitri and Ishiguro, Akio and Ijspeert, Auke J.},
doi = {10.1126/scirobotics.abf6354},
issn = {24709476},
journal = {Science Robotics},
number = {57},
pmid = {34380756},
title = {{Emergence of robust self-organized undulatory swimming based on local hydrodynamic force sensing}},
volume = {6},
year = {2021}
}

@article{Li2021,
author = {Li, Liang and Ravi, Sridhar and Xie, Guangming and Couzin, Iain D.},
doi = {10.1098/rspa.2020.0810},
issn = {14712946},
journal = {Proceedings of the Royal Society A: Mathematical, Physical and Engineering Sciences},
number = {2249},
title = {{Using a robotic platform to study the influence of relative tailbeat phase on the energetic costs of side-by-side swimming in fish}},
volume = {477},
year = {2021}
}

@article{Li2020,
author = {Li, Liang and Nagy, M{\'{a}}t{\'{e}} and Graving, Jacob M. and Bak-Coleman, Joseph and Xie, Guangming and Couzin, Iain D.},
doi = {10.1038/s41467-020-19086-0},
issn = {20411723},
journal = {Nature Communications},
number = {1},
pages = {1--9},
pmid = {33106484},
publisher = {Springer US},
title = {{Vortex phase matching as a strategy for schooling in robots and in fish}},
url = {http://dx.doi.org/10.1038/s41467-020-19086-0},
volume = {11},
year = {2020}
}

@article{Porez2014,
author = {Porez, Mathieu and Boyer, Fr{\'{e}}d{\'{e}}ric and Ijspeert, Auke Jan},
doi = {10.1177/0278364914525811},
issn = {17413176},
journal = {International Journal of Robotics Research},
month = {sep},
number = {10},
pages = {1322--1341},
publisher = {SAGE Publications Inc.},
title = {{Improved lighthill fish swimming model for bio-inspired robots: Modeling, computational aspects and experimental comparisons}},
volume = {33},
year = {2014}
}

@article{Manfredi2013,
author = {Manfredi, L. and Assaf, T. and Mintchev, S. and Marrazza, S. and Capantini, L. and Orofino, S. and Ascari, L. and Grillner, S. and Wall{\'{e}}n, P. and Ekeberg, {\"{O}} and Stefanini, C. and Dario, P.},
doi = {10.1007/s00422-013-0566-2},
issn = {03401200},
journal = {Biological Cybernetics},
number = {5},
pages = {513--527},
pmid = {24030051},
title = {{A bioinspired autonomous swimming robot as a tool for studying goal-directed locomotion}},
volume = {107},
year = {2013}
}

@article{Yan2008,
author = {Yan, Qin and Han, Zhen and wu Zhang, Shi and Yang, Jie},
doi = {10.1016/S1672-6529(08)60012-8},
issn = {16726529},
journal = {Journal of Bionic Engineering},
number = {2},
pages = {95--101},
title = {{Parametric Research of Experiments on a Carangiform Robotic Fish}},
volume = {5},
year = {2008}
}

@article{Anastasiadis2023,
author = {Anastasiadis, Alexandros and Paez, Laura and Tytell, Eric and Ijspeert, Auke Jan and Mulleners, Karen},
doi = {10.1038/s41598-023-41074-9},
isbn = {4159802341074},
issn = {2045-2322},
journal = {Scientific Reports},
pages = {1--12},
publisher = {Nature Publishing Group UK},
title = {{Identification of the trade-off between speed and efficiency in undulatory swimming using a bio-inspired robot}},
url = {https://doi.org/10.1038/s41598-023-41074-9},
year = {2023}
}

@article{Anastasiadis2024,
author = {Anastasiadis, Alexandros and Rossi, Annalisa and Paez, Laura and Melo, Kamilo and Tytell, Eric D and Ijspeert, Auke J and Mulleners, Karen},
doi = {10.1103/PhysRevFluids.9.110509},
journal = {Physical Review Fluids},
pages = {3--6},
publisher = {American Physical Society},
title = {{Eel-like robot swims more efficiently with increasing joint amplitudes compared to constant joint amplitudes}},
volume = {110509},
year = {2024}
}

@article{Bayat2016,
author = {Bayat, Behzad and Crespi, Alessandro and Ijspeert, Auke},
doi = {10.1109/AUV.2016.7778700},
isbn = {9781509024421},
journal = {Autonomous Underwater Vehicles 2016, AUV 2016},
pages = {381--386},
publisher = {IEEE},
title = {{Envirobot: A Bio-inspired environmental monitoring platform}},
year = {2016}
}

@article{Santo2021,
	author = {Santo, Valentina Di and Goerig, Elsa and Wainwright, Dylan K. and Akanyeti, Otar and Liao, James C. and Castro-Santos, Theodore and Lauder, George V.},
	doi = {10.1073/pnas.2113206118},
	issn = {10916490},
	journal = {Proceedings of the National Academy of Sciences of the United States of America},
	number = {49},
	pages = {1--9},
	pmid = {34853171},
	title = {{Convergence of undulatory swimming kinematics across a diversity of fishes}},
	volume = {118},
	year = {2021}
}

@article{Raynaud2025,
	author = {Raynaud, Ga{\'{e}}tan and Mulleners, Karen},
	doi = {10.1088/1361-6501/adc1e7},
	issn = {13616501},
	journal = {Measurement Science and Technology},
	number = {4},
	title = {{Event-based reconstruction of time-resolved centreline deformation of flapping flags}},
	volume = {36},
	year = {2025}
}

@article{Eidietis2002,
	author = {Eidietis, L. and Forrester, T. L. and Webb, P. W.},
	doi = {10.1139/z02-203},
	issn = {00084301},
	journal = {Canadian Journal of Zoology},
	number = {12},
	pages = {2156--2163},
	title = {{Relative abilities to correct rolling disturbances of three morphologically different fish}},
	volume = {80},
	year = {2002}
}

@article{Webb2002,
	author = {Webb, Paul W.},
	doi = {10.1093/icb/42.1.94},
	issn = {00031569},
	journal = {Integrative and Comparative Biology},
	number = {1},
	pages = {94--101},
	title = {{Control of posture, depth, and swimming trajectories of fishes}},
	volume = {42},
	year = {2002}
}

@article{Shepard2010,
	author = {Shepard, Emily L.C. and Wilson, Rory P. and Quintana, Flavio and Laich, Agustina G{\'{o}}mez and Liebsch, Nikolai and Albareda, Diego A. and Halsey, Lewis G. and Gleiss, Adrian and Morgan, David T. and Myers, Andrew E. and Newman, Chris and Macdonald, David W.},
	doi = {10.3354/esr00084},
	issn = {18635407},
	journal = {Endangered Species Research},
	number = {1},
	pages = {47--60},
	title = {{Identification of animal movement patterns using tri-axial accelerometry}},
	volume = {10},
	year = {2010}
}

@INPROCEEDINGS{Xie2023,
	author={Xie, Xiao and Herault, Johann and Lebastard, Vincent and Boyer, Frédéric},
	booktitle={2023 IEEE International Conference on Advanced Robotics and Its Social Impacts (ARSO)}, 
	title={Recursive inverse dynamics of a swimming snake-like robot with a tree-like mechanical structure}, 
	year={2023},
	volume={},
	number={},
	pages={65-70},
	doi={10.1109/ARSO56563.2023.10187577}
}

@article{White2021,
	author = {White, Carl H. and Lauder, George V. and Bart-Smith, Hilary},
	doi = {10.1088/1748-3190/abb86d},
	issn = {17483190},
	journal = {Bioinspiration and Biomimetics},
	number = {2},
	publisher = {IOP Publishing},
	title = {{Tunabot Flex: A tuna-inspired robot with body flexibility improves high-performance swimming}},
	volume = {16},
	year = {2021}
}

@inproceedings{Herault2020,
	address = {Cham},
	author = {Herault, Johann and Clement, {\'{E}}tienne and Brossillon, Jonathan and LaGrange, Seth and Lebastard, Vincent and Boyer, Frederic},
	booktitle = {Biomimetic and Biohybrid Systems},
	editor = {Vouloutsi, Vasiliki and Mura, Anna and Tauber, Falk and Speck, Thomas and Prescott, Tony J and Verschure, Paul F M J},
	isbn = {978-3-030-64313-3},
	pages = {165--175},
	publisher = {Springer International Publishing},
	title = {{Standing on the Water: Stability Mechanisms of Snakes on Free Surface}},
	year = {2020}
}

@article{Katzschmann2018,
	author = {Katzschmann, Robert K. and DelPreto, Joseph and MacCurdy, Robert and Rus, Daniela},
	doi = {10.1126/SCIROBOTICS.AAR3449},
	issn = {24709476},
	journal = {Science Robotics},
	number = {16},
	pages = {1--13},
	pmid = {33141748},
	title = {{Exploration of underwater life with an acoustically controlled soft robotic fish}},
	volume = {3},
	year = {2018}
}

@article{Tytell2004a,
	author = {Tytell, Eric D.},
	doi = {10.1242/jeb.01139},
	issn = {00220949},
	journal = {Journal of Experimental Biology},
	number = {19},
	pages = {3265--3279},
	pmid = {15326203},
	title = {{The hydrodynamics of eel swimming II. Effect of swimming speed}},
	volume = {207},
	year = {2004}
}

@article{Luczak2017,
author = {Łuczak, Sergiusz and Grepl, Robert and Bodnicki, Maciej},
title = {Selection of MEMS Accelerometers for Tilt Measurements},
journal = {Journal of Sensors},
volume = {2017},
number = {1},
pages = {9796146},
doi = {https://doi.org/10.1155/2017/9796146},
url = {https://onlinelibrary.wiley.com/doi/abs/10.1155/2017/9796146},
eprint = {https://onlinelibrary.wiley.com/doi/pdf/10.1155/2017/9796146},
year = {2017}
}

@article{Huang2016,
	author = {Huang, Hen Wei and Sakar, Mahmut Selman and Petruska, Andrew J. and Pan{\'{e}}, Salvador and Nelson, Bradley J.},
	doi = {10.1038/ncomms12263},
	issn = {20411723},
	journal = {Nature Communications},
	pages = {1--10},
	pmid = {27447088},
	title = {{Soft micromachines with programmable motility and morphology}},
	volume = {7},
	year = {2016}
}

@article{Gray1933,
    author = {Gray, J.},
	title = {Studies in Animal Locomotion: I. The Movement of Fish with Special Reference to the Eel},
	journal = {Journal of Experimental Biology},
	volume = {10},
	number = {1},
	pages = {88-104},
	year = {1933},
	month = {01},
	issn = {0022-0949},
	doi = {10.1242/jeb.10.1.88},
	url = {https://doi.org/10.1242/jeb.10.1.88},
	eprint = {https://journals.biologists.com/jeb/article-pdf/10/1/88/2610299/jexbio_10_1_88.pdf},	
}

@article{Lighthill1970,
	author = {Lighthill, M. J.},
	doi = {10.1017/S0022112070001830},
	issn = {14697645},
	journal = {Journal of Fluid Mechanics},
	number = {2},
	pages = {265--301},
	title = {{Aquatic animal propulsion of high hydromechanical efficiency}},
	volume = {44},
	year = {1970}
}

@article{Epps2007,
	author = {Epps, Brenden P. and Techet, Alexandra H.},
	doi = {10.1007/s00348-007-0401-4},
	issn = {07234864},
	journal = {Experiments in Fluids},
	number = {5},
	pages = {691--700},
	title = {{Impulse generated during unsteady maneuvering of swimming fish}},
	volume = {43},
	year = {2007}
}

@article{Triantafyllou2016,
	author = {Triantafyllou, Michael S. and Weymouth, Gabriel D. and Miao, Jianmin},
	doi = {10.1146/annurev-fluid-122414-034329},
	issn = {00664189},
	journal = {Annual Review of Fluid Mechanics},
	number = {July 2015},
	pages = {1--24},
	title = {{Biomimetic Survival Hydrodynamics and Flow Sensing}},
	volume = {48},
	year = {2016}
}

@article{King2023,
	author = {King, Justin T. and Green, Melissa A.},
	doi = {10.1007/s00348-023-03655-2},
	isbn = {0123456789},
	issn = {14321114},
	journal = {Experiments in Fluids},
	number = {6},
	pages = {1--17},
	publisher = {Springer Berlin Heidelberg},
	title = {{The influence of trailing edge shape on the wake circulation and time-averaged wake of bio-inspired pitching panels}},
	url = {https://doi.org/10.1007/s00348-023-03655-2},
	volume = {64},
	year = {2023}
}

@article{Zhu2019,
	author = {Zhu, J. and White, C. and Wainwright, D. K. and {Di Santo}, V. and Lauder, G. V. and Bart-Smith, H.},
	doi = {10.1126/scirobotics.aax4615},
	issn = {24709476},
	journal = {Science Robotics},
	number = {34},
	pmid = {33137777},
	title = {{Tuna robotics: A high-frequency experimental platform exploring the performance space of swimming fishes}},
	volume = {4},
	year = {2019}
}

@article{Wu2017,
	author={Wu, Zhengxing and Liu, Jincun and Yu, Junzhi and Fang, Hao},
	journal={IEEE/ASME Transactions on Mechatronics}, 
	title={Development of a Novel Robotic Dolphin and Its Application to Water Quality Monitoring}, 
	year={2017},
	volume={22},
	number={5},
	pages={2130-2140},
	doi={10.1109/TMECH.2017.2722009}}

@article{Wainwright2020,
		author = {Wainwright, Dylan K. and Lauder, George V.},
		doi = {10.1088/1748-3190/ab75f7},
		issn = {17483190},
		journal = {Bioinspiration and Biomimetics},
		number = {3},
		pmid = {32053798},
		publisher = {IOP Publishing},
		title = {{Tunas as a high-performance fish platform for inspiring the next generation of autonomous underwater vehicles}},
		volume = {15},
		year = {2020}
	}

@article{Lauder2007,
		author = {Lauder, George V. and Anderson, Erik J. and Tangorra, James and Madden, Peter G.A.},
		doi = {10.1242/jeb.000265},
		issn = {00220949},
		journal = {Journal of Experimental Biology},
		number = {16},
		pages = {2767--2780},
		pmid = {17690224},
		title = {{Fish biorobotics: Kinematics and hydrodynamics of self-propulsion}},
		volume = {210},
		year = {2007}
	}

@article{Bhushan2009,
		author = {Bhushan, Bharat },
		title = {Biomimetics: lessons from nature–an overview},
		journal = {Philosophical Transactions of the Royal Society A: Mathematical, Physical and Engineering Sciences},
		volume = {367},
		number = {1893},
		pages = {1445-1486},
		year = {2009},
		doi = {10.1098/rsta.2009.0011},
		
		URL = {https://royalsocietypublishing.org/doi/abs/10.1098/rsta.2009.0011},
		eprint = {https://royalsocietypublishing.org/doi/pdf/10.1098/rsta.2009.0011}
	}

@article{Smits2019,
		author = {Smits, Alexander J.},
		doi = {10.1017/jfm.2019.284},
		issn = {14697645},
		journal = {Journal of Fluid Mechanics},
		title = {{Undulatory and oscillatory swimming}},
		year = {2019}
	}

@article{Cui2023,
		author = {Cui, Zhongao and Li, Liao and Wang, Yuhang and Zhong, Zhiwei and Li, Junyang},
		doi = {10.1007/s44295-023-00010-3},
		issn = {29481953},
		journal = {Intelligent Marine Technology and Systems},
		number = {1},
		pages = {1--28},
		publisher = {Springer Nature Singapore},
		title = {{Review of research and control technology of underwater bionic robots}},
		url = {https://doi.org/10.1007/s44295-023-00010-3},
		volume = {1},
		year = {2023}
	}

@article{Ramdya2023,
		author = {Pavan Ramdya  and Auke Jan Ijspeert },
		title = {The neuromechanics of animal locomotion: From biology to robotics and back},
		journal = {Science Robotics},
		volume = {8},
		number = {78},
		pages = {eadg0279},
		year = {2023},
		doi = {10.1126/scirobotics.adg0279},
		URL = {https://www.science.org/doi/abs/10.1126/scirobotics.adg0279},
		eprint = {https://www.science.org/doi/pdf/10.1126/scirobotics.adg0279}}

@article{VanGinneken2005,
			author = {{Van Ginneken}, Vincent and Antonissen, Erik and M{\"{u}}ller, Ulrike K. and Booms, Ronald and Eding, Ep and Verreth, Johan and {Van Den Thillart}, Guido},
			doi = {10.1242/jeb.01524},
			issn = {00220949},
			journal = {Journal of Experimental Biology},
			number = {7},
			pages = {1329--1335},
			pmid = {15781893},
			title = {{Eel migration to the Sargasso: Remarkably high swimming efficiency and low energy costs}},
			volume = {208},
			year = {2005}
		}

@article{Gazzola2014,
			author = {Gazzola, Mattia and Argentina, M{\'{e}}d{\'{e}}ric and Mahadevan, L.},
			doi = {10.1038/nphys3078},
			issn = {17452481},
			journal = {Nature Physics},
			number = {10},
			pages = {758--761},
			title = {{Scaling macroscopic aquatic locomotion}},
			volume = {10},
			year = {2014}
		}

@article{Sverdrup-Thygeson2016,
		author = {Sverdrup-Thygeson, J. and Kelasidi, E. and Pettersen, K. Y. and Gravdahl, J. T.},
		doi = {10.1109/AUV.2016.7778701},
		isbn = {9781509024421},
		journal = {Autonomous Underwater Vehicles 2016, AUV 2016},
		pages = {387--395},
		publisher = {IEEE},
		title = {{The underwater swimming manipulator - A bio-inspired AUV}},
		year = {2016}
	}

@article{Triantafyllou2000,
		author = {Triantafyllou, Michael S. and Triantafyllou, G. S. and Yue, D. K. P.},
		issn = {0066-4189},
		journal = {Annual Review of Fluid Mechanics},
		pages = {33--53},
		title = {{Ydrodynamics of Fishlike Swimming}},
		volume = {32},
		year = {2000}
	}

@article{Gravish2018,
		author = {Gravish, Nick and Lauder, George V.},
		doi = {10.1242/jeb.138438},
		issn = {00220949},
		journal = {Journal of Experimental Biology},
		number = {7},
		pages = {1--8},
		pmid = {29599417},
		title = {{Robotics-inspired biology}},
		volume = {221},
		year = {2018}
	}

@article{Leftwich2012,
			author = {Leftwich, Megan C. and Tytell, Eric D. and Cohen, Avis H. and Smits, Alexander J.},
			doi = {10.1242/jeb.061440},
			issn = {00220949},
			journal = {Journal of Experimental Biology},
			number = {3},
			pages = {416--425},
			pmid = {22246250},
			title = {{Wake structures behind a swimming robotic lamprey with a passively flexible tail}},
			volume = {215},
			year = {2012}
		}

@article{Quinn2015,
			author = {Quinn, Daniel B. and Lauder, George V. and Smits, Alexander J.},
			doi = {10.1017/jfm.2015.35},
			issn = {14697645},
			journal = {Journal of Fluid Mechanics},
			pages = {430--448},
			title = {{Maximizing the efficiency of a flexible propulsor using experimental optimization}},
			volume = {767},
			year = {2015}
		}

@article{Webb1976,
		    author = {Webb, P. W.},
		title = {The Effect of Size on the Fast-Start Performance of Rainbow Trout Salmo Gairdneri, and A Consideration of Piscivorous Predator-Prey Interactions},
		journal = {Journal of Experimental Biology},
		volume = {65},
		number = {1},
		pages = {157-178},
		year = {1976},
		month = {08},
		issn = {0022-0949},
		doi = {10.1242/jeb.65.1.157},
		url = {https://doi.org/10.1242/jeb.65.1.157},
		eprint = {https://journals.biologists.com/jeb/article-pdf/65/1/157/3186956/jexbio_65_1_157.pdf},
		}

@article{Bainbridge1958,
		author = {Bainbridge, Richard},
		doi = {10.1242/jeb.35.1.109},
		issn = {0022-0949},
		journal = {Journal of Experimental Biology},
		number = {1},
		pages = {109--133},
		title = {{The Speed of Swimming of Fish as Related to Size and to the Frequency and Amplitude of the Tail Beat}},
		volume = {35},
		year = {1958}
	}

@article{Fish2020,
		author = {Fish, Frank E.},
		doi = {10.1088/1748-3190/ab5a34},
		issn = {17483190},
		journal = {Bioinspiration and Biomimetics},
		number = {2},
		pmid = {31751980},
		publisher = {IOP Publishing},
		title = {{Advantages of aquatic animals as models for bio-inspired drones over present AUV technology}},
		volume = {15},
		year = {2020}
	}

@article{Gemmell2015,
		author = {Gemmell, Brad J. and Colin, Sean P. and Costello, John H. and Dabiri, John O.},
		doi = {10.1038/ncomms9790},
		issn = {20411723},
		journal = {Nature Communications},
		pages = {1--8},
		pmid = {26529342},
		publisher = {Nature Publishing Group},
		title = {{Suction-based propulsion as a basis for efficient animal swimming}},
		volume = {6},
		year = {2015}
	}

@article{Nangia2017,
		author = {Nangia, Nishant and Bale, Rahul and Chen, Nelson and Hanna, Yohanna and Patankar, Neelesh A.},
		doi = {10.1371/journal.pone.0179727},
		isbn = {1111111111},
		issn = {19326203},
		journal = {PLoS ONE},
		number = {6},
		pages = {1--23},
		pmid = {28654649},
		title = {{Optimal specific wavelength for maximum thrust production in undulatory propulsion}},
		volume = {12},
		year = {2017}
	}

@article{Bayat2017,
		author = {Bayat, Behzad and Crasta, Naveena and Crespi, Alessandro and Pascoal, Ant{\'{o}}nio M. and Ijspeert, Auke J},
		doi = {10.1016/j.copbio.2017.01.009},
		issn = {18790429},
		journal = {Current Opinion in Biotechnology},
		number = {645141},
		pages = {76--84},
		pmid = {28254670},
		title = {{Environmental monitoring using autonomous vehicles: a survey of recent searching techniques}},
		volume = {45},
		year = {2017}
	}

@article{Berlinger2021,
		author = {Berlinger, F. and Saadat, M. and Haj-Hariri, H. and Lauder, George V. and Nagpal, R.},
		doi = {10.1088/1748-3190/abd013},
		issn = {17483190},
		journal = {Bioinspiration and Biomimetics},
		number = {2},
		publisher = {IOP Publishing},
		title = {{Fish-like three-dimensional swimming with an autonomous, multi-fin, and biomimetic robot}},
		volume = {16},
		year = {2021}
	}

@article{Liu2025,
	author = {Xiangxiao Liu  and Matthew D. Loring  and Luca Zunino  and Kaitlyn E. Fouke  and François A. Longchamp  and Alexandre Bernardino  and Auke J. Ijspeert  and Eva A. Naumann },
	title = {Artificial embodied circuits uncover neural architectures of vertebrate visuomotor behaviors},
	journal = {Science Robotics},
	volume = {10},
	number = {107},
	pages = {eadv4408},
	year = {2025},
	doi = {10.1126/scirobotics.adv4408},
	URL = {https://www.science.org/doi/abs/10.1126/scirobotics.adv4408},
	eprint = {https://www.science.org/doi/pdf/10.1126/scirobotics.adv4408}}

\end{document}